\documentclass[12pt,a4paper]{article}
\usepackage[T2A]{fontenc}
\usepackage[cp866]{inputenc}
\usepackage[english]{babel}
\usepackage[dvips]{graphicx }
\usepackage{axodraw,epsfig}
\usepackage{amsmath,euscript,amssymb}
\newcommand{\be}{\begin{equation}}
\newcommand{\ee}{\end{equation}}
\newcommand{\bea}{\begin{eqnarray}}
\newcommand{\eea}{\end{eqnarray}}

\title{Instantaneous Interactions in Standard Model}
\medskip
\author{A.Z. Dubni\v{c}kov\'{a}$^{1},$ S. Dubni\v{c}ka$^{2},$
 V.N. Pervushin$^{3}$, M. Se\v cansk\'y$^{2,3}$ \\
\\
$^{1}${\small Department of Theoretical Physics, Comenius
University, Bratislava, Slovak Republic}\\
$^{2}${\small  Institute of Physics, Slovak Academy of Sciences,
Bratislava, Slovak Republic}\\$^{3}${\small Joint Institute for
Nuclear Research, 141980 Dubna, Russia} }

\begin{document}

\maketitle

\begin{abstract}

 {Instantaneous interactions in Standard Model
 are derived by the resolution
 of the Gauss constraints in  inertial reference frames
 defined as  irreducible
  representations of the Poincar\'e group.
 The relativistic covariant and gauge-invariant S-matrix
 is constructed with taking into account
 instantaneous interactions.
     Physical effects formed by
 the instantaneous interactions  are discussed,
 including
 the  enhancement in the
 $K \to \pi$ transition probability and
 the nonleptonic and semileptonic kaon decays. The low-energy
 relations between these decay probability amplitudes and meson form factors
  are established. Exploiting these relations we estimate
 possible parameters of
   the differential $K \to \pi e^+ e^-$ decay
  rate in the NA48/2 CERN experiment.}

\end{abstract}

\vspace{3cm}

 \centerline{Invited talk  at}

 \centerline{The XVIII International Baldin Seminar on High Energy Physics}

 \centerline{{\it Problems of Relativistic Nuclear Physics
 and Quantum Chromodynamics}}

  \centerline{Joint Institute for Nuclear Research,
    Dubna, Russia, 25 -- 30 September, 2006}
\newpage


\newpage

\section{Introduction}

 The instantaneous Coulomb interaction in QED is well
 known as a fixed experimental
 fact.
 Instantaneous interaction  is one of the indispensable attributes of
 the Dirac Hamiltonian approach  to QED \cite{d,pol},
 where a variable  without the  canonical momentum is converted  into
  a  Coulomb ``potential''.
   Remember that this Hamiltonian approach was
   used   \cite{f1,pvn3} as
  the foundation of all heuristic methods of quantization
  of gauge theories, including the Faddeev-Popov (FP) method
  \cite{fp1} and
   all its   versions used now for description of
   Standard Model of elementary
   particles.

 However, a contemporary reader could not find
 instantaneous interactions in the accepted version of
 Standard Model (SM) of elementary particles  based on
 the FP  method \cite{db}.

 Why did the accepted version of SM  omit instantaneous interactions?

  What are the fundamental principles of foundation
 of instantaneous interactions?

 What are physical results following from the instantaneous
 interactions omitted by the  accepted version of SM?

 Responses to these questions
 are the topics of the present paper.

 In Section 2, we recall the Dirac Hamiltonian
 approach to gauge theories, where the  instantaneous interactions
 was introduced by Dirac \cite{d} in agreement with the priority of
  quantum principles and experimental facts.
 Section 3 is devoted to the statement of the problem
 of description of instantaneous interactions in SM.
 In Section 4,  some
 experimental tests are discussed as evidences of instantaneous interactions.
 In Appendix {\bf A}, S-matrix for bound states in electrodynamics
 is derived in terms of bilocal fields. In Appendix {\bf B},
   the problem of the
  chiral hadronization of QCD in the spirit  of
  the non-Abelian generalization of the Dirac approach to QED
   is discussed.


\section{Status of instantaneous interaction in gauge theories}

\subsection{The frame-depended variational principle}

 In order to explain the origin of the instantaneous interaction in SM,
 let us consider the Dirac approach \cite{d} to QED given by
 the action
 \be\label{1e} S_{\rm QED}=\int d^4x{\cal L}_{\rm QED},
 \ee
 with Lagrangian
 \be\label{2e}
 {\cal L}_{\rm QED}=
 -\frac{1}{4}\left[\partial_\mu A_\nu -\partial_\nu A_\mu\right]^2+
 \bar \psi [i\rlap/\partial -m]\psi-A_\mu j^{\,\mu}.
 \ee
 where $\rlap/\partial = \partial^{\mu} \gamma_{\mu}$,
 $A_{\mu}$ is a vector potential, $\psi$ is the
 electron-positron field described by the Dirac bispinor,
 $j_{\mu}=e  \bar {\psi}  \gamma_{\mu} \psi$ is the charge current,
 and $e$ is the electron charge.

 The action (\ref{1e}) is invariant with respect to gauge
 transformations \cite{fock,wh}
 \bea
 \label{3e1}
 A^{\lambda}_\mu=A_\mu+\partial_\mu\lambda,~~~~
 \psi^{\lambda}=e^{+\imath e\lambda}\psi,
 ~~~~~\bar \psi^{\lambda}=e^{-\imath e\lambda}\bar \psi.
 \eea

 Physical solutions of the system of variational equations
 \be\label{vp}\frac{\delta S_{\rm QED}}{\delta
 A_\mu}=\partial_\nu \left[\partial^\nu A^\mu -
 \partial^\mu A^\nu\right]-j^\mu=0
 \ee
 are obtained in a specific {\it inertial frame of
 reference to initial data}\footnote{The physical concept of
 a {\it frame of
 reference to initial data} is defined
 as  a three-dimensional coordinate basis with a watch and
 a ruler (and
 other physical devices)
 for measurement of time and distance (velocity, mass, and other
physical quantities), i.e. the {\it initial data} required for
unambiguous resolving equations of motion (\ref{vp}). {\it
Inertial} means that this  coordinate basis is connected with a
heavy physical body  moving  without influences of any  external
forces. Inertial frame of reference in Minkowskian space-time is
associated with the unit time-axis $n_\mu=
 (\frac{1}{\sqrt{1-\vec v^2}},\frac{\vec v}{\sqrt{1-\vec v^2}})$,
 where $\vec v$ is the velocity.
 The frame of reference $n^{\rm cf}_\mu=(1,0,0,0)$ with $\vec v=0$
 is called the comoving frame.
 Transition to another {\it inertial frame} are fulfilled by
a Lorentz transformation $L_{\mu\nu}n^{\rm cf}_\mu=n_\mu$.
 A complete set of inertial frames $\{n_\mu\}$ is
 obtained by all Lorentz transformations
 of  a comoving frame.
\label{f-1}}
  distinguished by a unit timelike vector $n_{\mu}$
 ($n^2_{\mu}=1$). This vector splits
 gauge field $A_\mu$ into the timelike $A_0=A_\mu n_{\mu}$
 and spacelike $A^{\bot}_\nu=A_\nu - n_{\nu}(A_\mu n_{\mu})$
 components\footnote{Recall that
 this specific reference frame was
 chosen by Wigner \cite{wigner}, in order to construct
  irreducible representations  of the Poincar\'e group
   supposing the existence of a vacuum as a state with
 minimal energy (see in detail \cite{61,Logunov69}).}.

 The time component of Eq. (\ref{vp}) with respect to
   $A_0$ is the Gauss constraint
\be\label{c1}
 \Delta A_0-\partial_0\partial_{k}A_k=-j_{0},
  \ee
 where $\Delta=\partial^2_{j}$, $j_0=e\bar \psi\gamma_0\psi$ is
 the  current.
 In accordance with the theory of differential equations,
  the field components $A_0$
  cannot be a {\it degree of freedom},
   because  the canonical momentum
 $P_0={\partial {\cal L}}/{\partial (\partial_0A_0)}=0
 $  is equal to zero.
 An exact solution
 of Eq. (\ref{c1})
\bea\label{c2}
  A_0(x_0, x_k)=\frac{1}{\Delta} [\partial_{k}~A_{k}-j_0]\equiv
  -\frac{1}{4\pi}\int d^3y
  \frac{[\partial_{k}~A_{k}(x_0,y_k)-j_0(x_0,y_k)]}{|\vec x-\vec y|}.
  \eea
 is treated as the {\it Coulomb potential field} leading to
 the {\it instantaneous} interaction. The
 source of this {\it potential field} can be only an
 electric  current $j_0$.
 The linear terms $\partial_0\partial_{k}A_k$ in this solution
  (\ref{c2}) cannot be considered as
 a physical source of the Coulomb potential. Therefore,
 Dirac \cite{d} supposed to remove this term from the Gauss
 constraint (\ref{c1})
 by gauge transformations
 \be\label{d1}
 A^{\rm (rad)}_\mu(A)=A_\mu+\partial_\mu\Lambda^{\rm (rad)}(A),~~
 ~~~\psi^{(\rm rad)}(\psi,A)=e^{\imath e\Lambda^{(\rm
 rad)}(A)}\psi,
 \ee
 where
  \bea\label{13-4}
  \Lambda^{(\rm rad)}(A)=-\frac{1}{\Delta} \partial_{k}~A_{k}\equiv
  \frac{1}{4\pi}\int d^3y
  \frac{\partial_{k}~A_{k}(x_0,y_k)}{|x-y|}.
  \eea

\subsection{Radiation variables as Dirac's  gauge-invariant observables}

This transformation  (\ref{d1}) determines new variables
\bea
 \label{13-1}
 A^{(\rm rad)}_0(A)&=&A_0+\partial_0\Lambda^{(\rm rad)}(A)=A_0
 -\partial_0\frac{1}{\Delta} \partial_{k}~A_{k},\\
 \label{13-2}
 A^{(\rm rad)}_l(A)&=&A_l+\partial_l\Lambda^{(\rm rad)}(A)=
 A_l-\partial_l\frac{1}{\Delta} \partial_{k}~A_{k},\\
 \label{13-3}
 \psi^{(\rm rad)}(\psi,A)&=&e^{\imath e\Lambda^{(\rm
 rad)}(A)}\psi.
 \eea
 Thus, the frame-fixing $A_\mu=(A_0,A_k)$, the
    treatment of $A_0$ as a classical field, and
  the Dirac  diagonalization  (\ref{d1}) of the Gauss law (\ref{c1}) \cite{d}
    lead to variables as
 the functionals that are invariant with respect to gauge transformations
  (\ref{3e1}) of the
 initial fields $A_\mu,\psi$
 \be\label{c3}
 A^{(\rm rad)}_\mu(A+\partial\lambda)=A^{(\rm rad)}_\mu(A),
 ~~~\psi^{(\rm rad)}(\psi e^{ie\lambda},A+\partial\lambda)=\psi^{(\rm
 rad)}(\psi,A).
 \ee
 These functionals were called the ``dressed fields'', or
 the ``radiation variables`` \cite{d}.

In terms of variables (\ref{13-1}) -- (\ref{13-3})
  the Gauss law  (\ref{c1}) takes the
 diagonal form (i.e. it loses the linear term $\partial_k A_k$)
 \be\label{1c1}
 \Delta A^{(\rm rad)}_0(A)=-j^{(\rm rad)}_{0}\equiv -e
 \bar \psi^{(\rm rad)}\gamma_0\psi^{(\rm rad)}
  \ee
 and the spatial components become transversal identically
\be\label{kc1}
 \partial_k A^{(\rm rad)}_k(A)\equiv {0}.
  \ee


 After the substitution of a manifest resolution
 of the Gauss constraint (\ref{c1})
  into the initial action (\ref{1e})\footnote{This substitution,
  i.e. the calculation of
  value of the action onto a solution of the Gauss
  constraint, is called the reduction procedure. This reduction
  allows us to eliminate nonphysical pure gauge degrees of
  freedom \cite{gpk}.}
   this  action
 can be expressed
 in terms of the gauge-invariant radiation variables
  (\ref{13-1}) -- (\ref{13-3})
 \cite{d,pol}
 \bea\label{14-2}
 &&S_{\rm QED}|_{\frac{\delta S_{\rm QED}}{\delta A_0}=0}
 \equiv S^{(\rm rad)}_{\rm QED}=\\
 \nonumber
  &&=
 \int d^4x \left[\frac{1}{2}(\partial_{\mu}A^{(\rm rad)}_k)^2
 +
 \bar \psi^{(\rm rad)}
 [i\rlap/\partial -m]\psi^{(\rm rad)}+
 A^{(\rm rad)}_kj^{(\rm rad)}_{k}-\frac{1}{2}
 j_0^{(\rm rad)}\frac{1}{\triangle}j_0^{(\rm rad)}\right]
 .
 \eea
 One can see that  Dirac in \cite{d} managed to  express his Hamiltonian
 scheme in terms of radiation variables  (\ref{13-1}) -- (\ref{13-3})
 as functionals
 from the initial fields, and these functionals  are invariant
(\ref{c3})
 with
 respect to  gauge transformations of the initial fields (\ref{3e1}).
In other words,
 one can say that
 the Dirac  radiation variables are manifestly gauge-invariant.
 Therefore, the fact that
 the  current $j^{(\rm rad)}_\mu$ cannot satisfy the
 conservation constraint
 \bea\label{14-3a}\partial_\mu {j^{(\rm rad)}}^\mu \not = 0.
 \eea
 does not contradict   the gauge-invariance of
 the initial action (\ref{1e}) and, moreover, it allows us to give
 the gauge-invariant description of
 bound states, where the  current $j^{(\rm rad)}_\mu$
  is not conserved.  

\subsection{Dirac's frame depended approach versus the  frame free  one}
 Values of the reduced action
 (\ref{14-2}) onto solutions of the motion equations takes the form
\bea\nonumber
 S^{(\rm rad)}_{\rm QED}|_{\frac{\delta S_{\rm QED}}{\delta A_k}=0}
 &=&\frac{1}{2}\int
 d^4xd^4y\left[
 j_0^{(\rm rad)}(x)\mbox{\rm \bf C}_{00}(x-y)j_0^{(\rm
 rad)}+j^{(\rm rad)}_{k}(x)\mbox{\rm \bf T}_{ki}(x-y)
 j^{(\rm rad)}_{i}(y)\right]+
\\
\label{15-1a}
  &+&
 \int d^4x \bar \psi^{(\rm rad)}
 [i\rlap/\partial -m]\psi^{(\rm rad)}
 ,
 \eea
 where
 \bea\label{15-2a}
 \mbox{\rm \bf C}_{00}(x-y)&=&-\delta(x_0-y_0)
 \frac{1}{4\pi|\vec{x}-\vec{y}|},\\\label{15-2b}
 \mbox{\rm \bf T}_{ki}(x-y)&=&\frac{1}{(2\pi)^4}\int
 d^4q\left(\delta_{ki}-
 \frac{q_kq_i}{{\vec{q}}^2}\right)\frac{e^{iq_\nu(x-y)^\nu}}{q_\mu^2+i\epsilon}.
 \eea
 The action (\ref{15-1a}) allows one to determine the
 propagator of radiation variables in the momentum representation
 $j(x)=(1/(2\pi)^2)\int d^4q e^{-iq_\nu x^\nu}\widetilde{j}(q)$.
 One can see that the  current part of the action (\ref{15-1a})
 takes the form
 \be\label{15-2}
 S^{{(\rm rad)}}[j]=\frac{1}{2}\int d^4q \widetilde{j}^\mu(q)
 \widetilde{j}^\nu(q)D^{\gamma,{(\rm rad)}}_{\mu\nu}(q),
 \ee
 where
 \be\label{wr}
  D^{\gamma,{(\rm rad)}}_{\mu\nu}(q) = \delta_{\mu
 0}\delta_{\nu 0}\frac{1}{\vec{q}^2} +\delta_{\mu i}\delta_{\nu j}
 \left(\delta_{ij}-\frac{q_iq_j}{\vec{q}^2}\right)\frac{1}{q_\mu^2+i\varepsilon}
 \equiv {\mbox{\rm \bf C}}_{\mu\nu}+{\mbox{\rm \bf T}}_{\mu\nu}\\
 \ee
 is the propagator as
    the sum of
 the instantaneous exchange ${\bf C}_{\mu\nu}$ and
 transverse field one ${\bf T}_{\mu\nu}$.

 This propagator contains two singularities:
  the Coulomb one  {\bf C} forming instantaneous atoms (see Appendix A) and
  the light cone ones  of the transverse variables
  {\bf T} (considered as radiation corrections).
 This propagator  can be identically rewritten as the sum of the
 Feynman propagator {\bf F}  (that does not depend on the frame
 time-axis $n_\mu$) and the longitudinal {\bf L} one:
 \bea
 \label{1wr}
 D^{\gamma,{(\rm rad)}}_{\mu\nu}(q) \equiv-\frac{\delta_{\mu \nu}}{q_\nu^2+i\varepsilon} +
 \frac{(q_0\delta_{\mu 0})(q_0\delta_{\nu 0})
 -(q_i\delta_{\mu i})(q_j\delta_{\nu j})}{\vec{q}^2[q_\nu^2+i\varepsilon]}
 \equiv {\mbox{\rm \bf F}}_{\mu\nu}(q)+{\mbox{\rm \bf L}}_{\mu\nu}(q).
 \eea

 It is easy to see that if the  currents $\widetilde{j}_\mu (q)$
 satisfy the conservation  constraint
 \bea\label{15-3}
 q^\mu \widetilde{j}^{(\rm rad)}_\mu  =0
 \eea
 in  contrast to  (\ref{14-3a}), the action (\ref{15-2}) takes the form
 \be\label{15-4}
 S^{{(\rm rad)}}[j]=\frac{1}{2}\int d^4q \widetilde{j}^\mu(q){\bf F}_{\mu\nu}(q)
 \widetilde{j}^\nu(q)
 \equiv-\frac{1}{2}\int d^4q
 \frac{\widetilde{j}^2_\mu(q)}{q_\nu^2+i\epsilon}.
 \ee
 This result corresponds to
 the Lagrangian
 \be\label{15-5}
 {\cal L}^{\rm F}_{\rm QED}={\cal L}_{\rm QED}-\frac{1}{2}(\partial_\mu
 A^\mu)^2,
 \ee
 where the first term is the initial Lagrangian (\ref{2e}).
 The change  
\be\label{15-6}
 {\cal L}_{\rm QED}|_{(\partial_\mu j^\mu=0)}~~\Rightarrow~~
 {\cal L}^{\rm F}_{\rm QED}={\cal L}_{\rm QED}-\frac{1}{2}(\partial_\mu
 A^\mu)^2
 \ee
  violates gauge-invariance (in contrast to the radiation variables)
  and it is known as the {\it Feynman gauge}.  A similar change (\ref{15-6}) is
  the basis of the accepted  frame free formulation
 of SM  \cite{db}.

 Thus, in  the Dirac Hamiltonian approach to QED we have perturbation
theory featuring two singularities in the photon propagators
(\ref{wr}) (the instantaneous singularity and that at the light
cone), whereas the  perturbation theory in the frame free gauge
(\ref{15-6}) (accepted now to formulate SM \cite{db}) involves
only one singularity in the propagators (that at the light cone
(\ref{15-4})). The perturbation theory in this gauge in terms of
the propagators  with the light cone singularity can by no means
describe instantaneous Coulomb atoms, and it contains only the
frame free Wick-Cutkosky bound states whose spectrum is not
observed in nature \cite{kummer,Nakanishi}.

 The problem of different physical results in two different {\it gauges}
  cannot be explained by the violation of the gauge-invariance
 in the Lagrangian  (\ref{15-6}). We can keep gauge-invariance
  of the frame free method choosing
 the Lorentz variables $A^{\rm (L)}_\mu(A)$ by gauge transformations
 \be\label{Ld1}
 A^{\rm (L)}_\mu(A)=A_\mu+\partial_\mu\Lambda^{\rm (L)}(A),
 ~~~\psi^{(\rm L)}(\psi,A)=e^{\imath e\Lambda^{(\rm
 L)}(A)}\psi, ~~~(\Lambda^{(\rm L)}(A)=-\frac{1}{\Box}
 \partial_{\mu}~A^\mu)~.
 \ee
 The gauge-invariant functional $A^{\rm (L)}_\mu(A)$
 identically satisfies the Lorentz constraint
 $ 
 \partial_\mu {A^{(\rm L)}}^\mu\equiv {0}.
  $ 



 In fact, the
 change from the radiation variable (\ref{d1}) $A^{(\rm rad)},\psi^{(\rm rad)}$
  to the frame free Lorentz ones (\ref{Ld1})
 $ A^{(\rm L)},\psi^{(\rm
 L)}$
 \bea
 \label{f13-1}
 A^{(\rm rad)}_\mu(A^{(\rm L)})&=&A^{(\rm L)}_\mu+
 \partial_\mu\Lambda^{(\rm rad)}(A^{(\rm L)}),\\
 \label{f13-3}
 \psi^{(\rm rad)}(\psi^{(\rm L)},A^{(\rm L)})&=&e^{\imath
 e\Lambda^{(\rm rad)}(A^{(\rm L)})}\psi^{(\rm L)},
 \eea
  where $\Lambda^{(\rm rad)}(A^{(\rm L)})$ is the Dirac phase given by Eq.
 (\ref{13-4}),  keeps all  physical results,
 if there remains the Dirac
 {\it gauge of physical sources}
 in terms of the new variables:
 \be\label{f16-10}
 {\cal L}^{(\rm rad)}_{\rm sources}= J^{(\rm rad)}_k [A^{(\rm
 L)}_k+\partial_k\Lambda^{(\rm rad)}(A^{(\rm L)})] +
  \bar \eta^{(\rm rad)}e^{\imath
 e\Lambda^{(\rm rad)}(A^{(\rm L)})}\psi^{(\rm L)}
  +\bar \psi^{(\rm L)}e^{-\imath
 e\Lambda^{(\rm rad)}(A^{(\rm L)})}\eta^{(\rm rad)}].
 \ee
  As a relic of the Dirac fundamental quantization, the Dirac phase
 factors $e^{\imath
 e\Lambda^{(\rm rad)}(A^{(\rm L)})}$ in the source Lagrangian
  (\ref{f16-10}) ``remember'' the entire body of
 information about the {\it frame of reference to initial data}
  and the instantaneous  interaction.
  As was
 predicted by Schwinger \cite{sch2}, all these effects disappear, leaving
 no trace, if these Dirac factors are removed by means of the
 substitution
\be\label{fl16-10}
 {\cal L}^{(\rm rad)}_{\rm sources}~~~~ \to ~~~~{\cal L}^{(\rm L)}_{\rm
 sources}
 = J^{(\rm L)}_\mu A^{(\rm
 L)}_\mu +
  \bar \eta^{(\rm L)}\psi^{(\rm L)}
  +\bar \psi^{(\rm L)}\eta^{(\rm L)}.
 \ee
 This substitution  (treated as another choice of {\it gauge of
 physical sources} by the elimination of the Dirac phase factors)
 loses  the instantaneous
 interaction and instantaneous bound states.
 We see that there is the strong
 dependence of
  physical results on the {\it gauge of physical sources},
 and this dependence  does not
  mean the violation of the gauge-invariance principle,
   because both the   radiation variables  (\ref{d1})  and
   the Lorentz ones  (\ref{Ld1}) 
   are gauge-invariant functionals of the initial
 fields\footnote{Therefore, the assertion that {\it ``the dependence
 of physical result on the choice of a {\it gauge}
 means the violation of the gauge-invariance principle''}
 is not correct.
 The correct full name of the concept
 ``gauge'' is {\it gauge of physical sources}.
 A change of {\it gauge of physical sources}
 can lead to a change of
  physical results as we have seen above.
}.

 The region of validity of the theorem equivalence of
 the frame dependent method
 and the frame free {\it Lorentz gauge formulations}
  (changing the {\it gauge of physical sources})
 is restricted by the elementary particle scattering
   amplitudes \cite{f1}.

 The change of the {\it gauge of physical sources} (\ref{fl16-10})
 is just the
 point where the frame free FP method in
 the {\it Lorentz gauge}  (\ref{15-6}) loses
 the instantaneous interaction (\ref{15-2a}) forming instantaneous
 bound states (considered in Appendixes A and B).
 Moreover,  the field $A_0$ treated as a dynamic one
 in the {\it the Lorentz gauge formulation} (\ref{15-6})
 gives a negative contribution to
 energy and it leads to tachyon in the spectrum of
 the Wick-Cutkosky  bound states \cite{kummer,Nakanishi}.

 The standard frame free approach \cite{db} to SM extending the FP method
  out of the region of its validity  (\ref{15-3}) finally   lost
  instantaneous interactions.
 However, if we keep instantaneous interactions,
 the  question arises about a choice of the frame and relativistic invariance.
 What do relativistic covariant
 instantaneous atoms  in QED  mean?

 \subsection{Relativistic invariance: ``frame free'' versus ``many frames''}

 It is interesting to trace the historical evolution of
 the physical concepts of {\it relativistic invariance}
 and {\it gauge invariance}.
 Both Julian Schwinger
 (who ``{\it  rejected all Lorentz gauge formulations
 as unsuited to the role of providing the fundamental operator
 quantization}'' \cite{sch2})
  and  Faddeev  (who accepted these
 {\it formulations} in \cite{f1}) used the same argument of {\it
 relativistic invariance}.
 The phrase
 ``{\it Finally, the Hamiltonian formulation, in the case of field theory,
   is not relativistically covariant}'' was treated in \cite{f1}
   as a defect of the frame dependent Hamiltonian formulation
     and one of the arguments in favor of the Lorentz frame free gauge
    (without the instantaneous interaction). However, in accord with
    the theory of representations of the Poincar\'e group
 the relativistic invariance
 means that {\it a complete set of states
 $\{|\Phi_I>\}_{n^{\rm cf}}$
 obtained by all Lorentz transformations
 of  a state $|\Phi>_0$
 in a definite inertial frame of reference $n^{\rm cf}_\mu=(1,0,0,0)$
 coincides with a complete set of states $\{|\Phi_I>\}_n$
 obtained by all Lorentz transformations of this state $|\Phi>_0$
 in another frame of reference} $n_\mu=
 (\frac{1}{\sqrt{1-\vec v^2}},\frac{\vec v}{\sqrt{1-\vec v^2}})$ \cite{61}.
 Therefore,
    the frame dependence of the Hamiltonian formulation
    was treated in \cite{sch2}
    as  necessity to prove that the algebra of commutation relations
    of gauge-invariant observables coincide with the Poincar\'e algebra.
 This proof was fulfilled
   by Zumino \cite{z} (and Schwinger \cite{sch2}
  in the case of non-Abelian fields even in 1962)
  at the level of the  algebra of commutation relations
  of the Hamiltonian and momentum operators
  with the nonlocal commutation relation
 $i[A^{(\rm rad)}_k(x),P^{(\rm rad)}_j(y)]=
 \left[\delta_{kj}-\frac{\partial_k\partial_j}
 {\triangle}\right]\delta^{3}(x-y)$
 for quantum radiation variables taken to be
  symmetrically ordering in  the Hamiltonian.

 We can see that the {\it relativistic invariance} as the theory of
irreducible representation of the Poincar\'e group does not
mean the frame free formulation (\ref{Ld1}).
 The {\it relativistic invariance} means that a
 complete set of all   physical  states includes the states obtained by
{\it all Lorentz transformations of the comoving frame},
i.e. all Lorentz transformations
 of the instantaneous interactions obtained
in the comoving frame. Thus,
 there are two following possibilities to construct
 the weak S-matrix amplitude transitions between
 physical states,
i) the frame free S-matrix, and ii) the many-frame S-matrix.

 \subsection{Many-frame relativistic  S-matrix with instantaneous interactions}

 Recall that S-matrix in QED is constructed on the basis of
 the Schr\"odinger equation
 \be\label{16-1}
  H^{(\rm rad)}|\Phi(x^0)>=
  i\frac{d}{dx^0}|\Phi(x^0)>,
 \ee
 where
 \bea\label{16-2}
  H^{(\rm rad)}&=&\int\limits_{V_0}^{}~d^3x~ {\cal H}^{(\rm rad)}[n]=
  \int\limits_{V_0}^{}~d^3x~ [{\cal H}_0(n)+{\cal H_{\rm ins.int.}}(n)+
  {\cal H_{\rm ret. int.}}(n)],\\\label{h16-2}
  {\cal H}^{(\rm rad)}(n)&=&[P_i^{(\rm rad)}\partial_0A_i^{(\rm rad)}+
  P_\psi^{(\rm rad)}
  \partial_0\psi^{(\rm rad)}-{\cal L}^{(\rm rad)}]
 \eea
 are the Hamiltonian and its density as
 a sum of
 free field part, instantaneous interaction  part,
 and retarded interaction one;  $\Phi$
 is a physical states.
  All these notions  (space-time, volume, energy, class of
  functions and etc.)
  make sense only in a definite {\it frame of
 reference to initial data}  distinguished by a
  unit time-axis $n_\mu=\left(\frac{1}{\sqrt{1-\vec v^2}},
  \frac{\vec v}{\sqrt{1-\vec v^2}}\right)$,
 where $\vec v$ is the initial velocity of an
 inertial frame, $\vec v=0$ is called  a {\it comoving frame}
 (cf) (see the footnote \ref{f-1}).
 The resolution of the equation takes the form of the
 operator of evolution
 \be\label{16-3}
 |\Phi(x^0)>=
  T\exp\left\{-i\int_{x^0_{(\rm in)}}^{x^0}
   dx^0H^{(\rm rad)}\right\}|\Phi_I(x^0_{\rm (in)})>,
 \ee
 here $|\Phi_I(x^0_{\rm (in)})>$ is
 an initial  state defined in the asymptotic region,
 where the {\it Hamiltonian  of retarded interaction} in the sum
 (\ref{16-2})
  can be neglected\footnote{We can neglect the retarded interaction
   provided that
  $
  |x^0_{\rm (out)}-x^0_{\rm (in)}|\gg E^{-1}_{I,\rm min},
  ~V_0^{1/3}\gg  E^{-1}_{I,\rm min}
  $.
 This condition means that all stationary
 solutions
 with  zero energy $E_{I,\rm min} \to 0$ cannot
 be considered as perturbational.},
 so that one can  determine  the energy spectrum $E_I$
  from the equation
 \be\label{16-5}
  \int d^3x[{\cal H}_0(n)+{\cal H}_{\rm ins.int.}(n)]|\Phi_I(x^0_{\rm (in)})>=
  E_I|\Phi_I(x^0_{\rm (in)})>.
 \ee
  In accord with the {\it spectrality principle}
   a complete set of eigenvalues $E_I$
 (with a finite energy density $E_I/V_0\leq\infty$)
 of a manifold of
  all asymptotic states with the quantum
 numbers $I$ includes a physical vacuum as a state with
 a minimal energy $E_{I,\rm min}$ \cite{Logunov69}.


 Notice that in the comoving frame $n^{\rm cf}_\mu=(1,0,0,0)
 $
 the set of one-particle and two-particle
 equations\footnote{Similar equations  are well - known from the
 nonrelativistic many - body theory (Landau's theory of fermi liquids
 \cite{fb24}, Random Phase Approximation \cite{fb25}) and play an essential
 role in the description of elementary excitation in atomic nuclei \cite{fb26}.
 Their relativistic analogies describe in QCD the Goldstone pion
 and the constituent masses of the light quarks \cite{5,6,7} (see Appendix B).} (\ref{16-5}) coincide with the Schwinger - Dyson equation
 and the Salpeter one, respectively (see Appendix A).

 The next step is the construction of the S-matrix elements as
 weak
 probability amplitude transitions between the all
asymptotic\footnote{The hypothesis of
 asymptotical states of S-matrix in QFT means that
   physical states are created together with their comoving
   {\it reference frames} where quantum numbers of these states are measured.}
 states including bound
 states.
  A complete set of these states includes
 any bound state moving with the momentum ${\cal P}_\mu$
 obtained the  Lorentz transformation of the time-axis
 \be\label{np}
 n^{\rm cf}_\mu=(1,0,0,0)~~\Longrightarrow ~~
 n^{\cal P}_\mu=\frac{{\cal P}_\mu}{\sqrt{{\cal P}^2}}=
 \left(\frac{{\cal P}_0}{\sqrt{{\cal P}^2}},~
 \frac{{\cal P}_k}{\sqrt{{\cal P}^2}}\right).
 \ee
Therefore,
   practically, in QED, one uses  the many-frame S-matrix
 \bea\label{w16-6}
 S^{(\rm rad, practical)}_{J_{\rm out},I_{\rm in}}=
  \left\langle\Phi_J(x^0_{\rm (out)})\left|
  \exp\left\{-i\int
   d^4x {\cal H}^{(\rm rad)}
   (\hat n^{\rm ID})\right\}\right|\Phi_I(x^0_{\rm
   (in)})\right\rangle,
 \eea
     where the
   time-axis $n_\mu^{\rm cf}=(1,0,0,0)$ \cite{sch2}
   is replaced by
   {\it   the initial datum operator} $\hat n^{\rm ID}_\mu$ proportional to the
  operator of the total
 momentum\footnote{It is clear that at the point of the
 existence of the bound state
 with the definite total momentum $ {\cal P}_{A\mu} $ any
 instantaneous interaction  with the time-axis $n_\mu$
 parallel to this momentum
 is much greater than any retarded interaction
  in the sum
 (\ref{16-2})~\cite{love}.}
   \be\label{3-al}
\hat n^{\rm ID}_\mu |\Phi^J>=\frac{P^J_\mu}{M_J}|\Phi^J>
   \ee
     of any physical state considered as
 a irreducible representation of the Poincar\'e group
   \cite{5,6}, where this operator  $\hat n^{\rm cf}_\mu$
 becomes c-number (see Appendix A).

 Thus, instead of the acceptable frame free formulation
 of the Standard Model \cite{db} without instantaneous interactions
 we suppose to use
  the Dirac-like radiation variables with instantaneous interactions
  and the many-frame S-matrix  (\ref{w16-6}) where the
 comoving time-axis is replaced by the initial datum operator
 (\ref{3-al})\footnote{In particular, this definition of
 S-matrix  (\ref{w16-6}) justifies the application of the axial gauge
 for the description of the deep-inelastic scattering amplitudes,
 where this gauge imitates the transition to the large momentum
frame \cite{feynman}.}.

  \section{Instantaneous weak interactions}
%

\subsection{Standard Model}

 We consider the accepted SM without any additional  terms
 violating gauge invariance
\begin{eqnarray}\nonumber
\mathcal{L}_{\phi}=\frac{1}{2}(\partial_{\mu}\phi)^{2}-
 \frac{\lambda}{4}[\phi^{2}-v^2]^2-\phi[f_v(\bar e_Re_L+\bar e_Le_R)+
 (\bar e_Re_L+\bar e_Le_R)]+\\\label{sm1}
+\frac{1}{8}\phi^{2}[g^{2}W_\mu^{+}W^{-\mu}
+(g^{2}+g'^{2})Z_{\mu}^{2}]
\end{eqnarray}
\begin{eqnarray}\nonumber
\mathcal{L}_{l}=\bar{\nu}_{e}i(\gamma\partial)\nu_{e}+\frac{g}{2\cos\theta_{W}}(\bar{\nu}_{e}\gamma^{L}_{\mu}\nu_{e})Z^{\mu}+
\bar{e}i\gamma^{\mu}(\partial_{\mu}+ig\sin\theta_{W}A_{\mu})e+\frac{g}{2\cos\theta_{W}}(\bar{e}\gamma^{L}_{\mu}e)Z^{\mu}+
\\\label{sm2}
+g\frac{\sin^{2}\theta_{W}}{\cos\theta_{W}}(\bar{e}\gamma_{\mu}e)Z^{\mu}+\frac{g}{\sqrt{2}}(\bar{\nu}_{e}W^{+}_{\mu}\gamma^{\mu}e_{L}+
\bar{e}_{L}W^{-}_{\mu}\gamma^{\mu}\nu_{e})
\end{eqnarray}
\begin{eqnarray}\label{sm3}
\!\!\!\mathcal{L}_{V}\!\!\!&=&\!\!\!-\frac{1}{4}(\partial_{\mu}A_{\nu}-\partial_{\nu}A_{\mu})^{2}
-\frac{1}{4}(\partial_{\mu}Z_{\nu}-\partial_{\nu}Z_{\mu})^{2}
-\frac{1}{2}[D_{\mu}W^{+}_{\nu}-D_{\nu}W^{+}_{\mu}]^{2}\\\nonumber
&-&ie(\partial_{\mu}A_{\nu}-\partial_{\nu}A_{\mu})W^{+\mu}W^{-\nu}
-g^{2}\cos^{2}\theta_{W}[Z_\nu^{2}(W_\mu^{+}W^{-\mu})-
(W_\mu^{+}Z^\mu)(W_\nu^{-}Z^\nu)] +\\\nonumber
\!\!\!&+&\!\!\!ig\cos_W\theta(\partial_{\mu}Z_{\nu}\!\!\!-\!\!\!\partial_{\nu}
Z_{\mu})W^{+\mu}W^{-\nu}
\!\!\!+\!\!\!\frac{1}{2}ig\cos_W\theta[(D_{\mu}W^{+}_{\nu}\!\!\!-\!\!\!
D_{\nu}W^{+}_{\mu})(W^{-\mu}Z^{\nu}\!\!\!-
\!\!\!W^{-\nu}Z^{\mu})\!\!\!-c.c.],
\end{eqnarray}
 where $V^K_\mu=(A_\mu,Z_\mu,W_\mu^{+},W_\mu^{-})$
 are the standard set of vector fields, and
 with the Higgs effect of spontaneous symmetry breaking
 \be\label{h-1}
 \phi=\eta +v, ~~~v^2=\frac{m^2_\eta}{2\lambda},
 \ee
 where $\eta$ is the Higgs field with the mass $m_\eta$.

 First of all, we can see that this action (\ref{sm3}) is bilinear
with
 respect to the time components of the vector fields $V^K_0=(A_0,Z_0,W_0^{+},W_0^{-})$
 in the ``comoving frame'' $n^{\rm cf}_\mu=(1,0,0,0)$
 \be\label{sm4}
 S_{V}=\int d^4x \left[\frac{1}{2}V^K_0\hat L^{KI}_{00}V^I_0+V^K_0J^K+...
 \right]~,
 \ee
  where $\hat L^{KI}_{00}$ is
 the matrix of differential operators. Therefore, the Dirac approach
to SM can be realized. This means that the
 problems of the reduction and
  diagonalization of the set of the Gauss laws are solvable, and
  the Poincar\'e algebra of gauge-invariant observables can be proved.

 In any case,  SM in the lowest order of  perturbation theory is reduced to
 the sum of the Abelian massive vector fields, where
  Dirac's  radiation variables was considered
 in \cite{hpp}.

\subsection{The reduction of the Abelian massive vector theory}

The classical action of massive QED is
\begin{equation}
\label{LQEDloc} W=\int d^4x {\cal L}(x)= \int d^4x
\left[-\frac{1}{4}F_{\mu\nu}F^{\mu\nu}+\frac{1}{2}M^2V_\mu^2+
\bar\Psi(i\rlap/\partial-m)\Psi -V_{\mu}J^{\mu}\right]~,
\end{equation}
with the  Lagrangian ${\cal L}(x)$ and the currents $J_\mu\equiv
e\bar\Psi\gamma_\mu\Psi$. The Euler-Lagrange equation for $V_0$ is
a constraint 
\begin{equation}\label{vgc}
\frac{\delta W}{\delta V_0}=0 \ \ \leftrightarrow \ \
(\triangle-M^2)V_0=-\partial_i\dot{V}_i+J_0~,
\end{equation}
 It has the
solution
\begin{equation}
\label{V_0V_iQED} V_0[\vec{V},J_0]
=\frac{1}{\triangle-M^2}(-\partial_i\dot{V}_i +J_0)~.
\end{equation}
Inserting this into W we obtain the reduced $W_{\rm red}$
\begin{equation}
W_{\rm red}[\vec{V},\Psi]=\int d^4x {\cal L}_{\rm red} \equiv \int
d^4x \left({\cal L}_{\rm red}^{V} +{\cal L}_{\rm red}^\Psi\right),
\label{WredmQED}
\end{equation}
where
\begin{eqnarray}
{\cal L}^V_{\rm red} &=&
\frac{1}{2}\left(\dot{V}_iR_{ij}\dot{V}_j+
                 V_i(\triangle-M^2)R_{ij}V_j\right)~,\nonumber\\
{\cal L}^\Psi_{\rm red} &=&
\frac{1}{2}J_0\frac{1}{\triangle-M^2}J_0
      +J_0\left(\frac{1}{\triangle-M^2}\partial_i\dot{V}_i\right)
                  +V_iJ_i+\bar\Psi(i\not{\!\partial}-m)\Psi~,
\label{Lredpsi}
\end{eqnarray}
are the reduced Lagrangians with the  operator
\begin{equation}
R_{ij}\equiv
\delta_{ij}-\frac{\partial_i\partial_j}{\triangle-M^2}=
\delta_{ij}^T-\frac{M^2}{\triangle-M^2}\delta_{ij}^{||}
\label{Projop}
\end{equation}
 where we have used the longitudinal and transverse projection
 operators
 \begin{equation} \label{projop} \delta_{ij}^{||}\equiv
\frac{\partial_i\partial_j}{\triangle}~,
 \ \ \ \ \ \ \
\delta_{ij}^T\equiv \delta_{ij}-\delta_{ij}^{||}~.
\end{equation}
 In contrast to the massless case,
$R_{ij}$ is not a projection operator,
 $R^2\neq R$, but
\begin{equation}
\label{R^2} R_{ij}R_{jl}=R_{il}+
\frac{M^2\partial_i\partial_l}{\left(\triangle-M^2\right)^2}
\end{equation}
and $R_{ij}$ is invertible
\begin{equation}
R_{ij}^{-1}=\delta_{ij}^T-\frac{\triangle-M^2}{M^2}\delta_{ij}^{||}=
\delta_{ij}-\frac{\partial_i\partial_j}{M^2}~.
\end{equation}
In the massless limit however the reduction operator becomes the
transverse projection operator for photons and ceases to be
invertible.


  The second term in ${\cal L}_{\rm
 red}^\Psi$ (\ref{Lredpsi}) can be eliminated (in order to diagonalize
 the Gauss law)
 by introducing the new radiation (R)
 variables \bea\label{1psired}
 V^{R}_j=\left(\delta_{ij}-\frac{\partial_i\partial_j}{\triangle-M^2}
      \right)V_j=R_{ij}V_j\\
 \label{psired}
 \Psi^{R}\equiv \exp\left(ie\frac{1}{\triangle-M^2}
                         \partial_iV_i\right)\Psi
 \eea
 which generalizes the  Dirac radiation
variables in QED  (\ref{13-2}) and  (\ref{13-3}). Since
(\ref{psired}) is only a phase transformation, the corresponding
current $J_{\mu}$ stays the same and ${\cal L}^\Psi_{\rm red}$
becomes
\begin{equation}
{\cal L}^\Psi_{\rm red} = \frac{1}{2}J_0\frac{1}{\triangle-M^2}J_0
      +J_i\left(\delta_{ij}-\frac{\partial_i\partial_j}{\triangle-M^2}
      \right)V_j +\bar\Psi^R(i\rlap/\partial-m)\Psi^R.
\end{equation}
Using the radiation variables $V^R_k=R_{kj}V_j$ with $R_{ij}$
given by (\ref{Projop}) the reduced Lagrangian can be written
\begin{eqnarray}
{\cal L}_{\rm red}= 
\frac{1}{2}\left(\dot{V}^R_iR^{-1}_{ij}\dot{V}^R_j+
               V^R_i(\triangle-M^2)R^{-1}_{ij}V^R_j\right)
             +\bar\Psi^R(i\rlap/\partial-m)\Psi^R +J_iV^R_i
                   +\frac{1}{2}J_0\frac{1}{\triangle-M^2}J_0
                  .
\label{redL}
\end{eqnarray}
Thus, the frame-fixing $A_\mu=(A_0,A_k)$, the
    treatment of $A_0$ as a classical field, and
  the Dirac-like  diagonalization of the Gauss law (\ref{vgc})
   (i.e. eliminating the linear field  contribution $\partial_i\dot{V}_i$)
    lead to radiation variables  (\ref{1psired}) and (\ref{psired})
     that cannot be associated with
    any gauge constraint,
    in the case of the massive fields \cite{hpp}.

\subsection{Dirac's frame dependent approach versus the frame free one}

The Fourier transform of the reduction operator $R_{ij}$
\begin{equation}
\sum_{\lambda}\epsilon_{i}^{P(\lambda)}(\vec{q})
\epsilon_{j}^{P(\lambda)}(\vec{q}) =
R_{ik}(q)\left(\delta_{kl}+\frac{q_k
q_l}{M^2}\right)R_{lj}(k)=R_{ij}(q)= \delta_{kl}-\frac{q_k
q_l}{\vec{q}^2+M^2}~.
\end{equation}
leads to the free propagator
\begin{eqnarray}
D^R_{ij}(x-y)&=&
\langle 0|TV^R_{i}(\vec{x},x_0)V^R_{j}(\vec{y},y_0)|0\rangle \nonumber\\
   &=&-i\int \frac{d^4q}{(2\pi)^4} \frac{e^{-iq\cdot (x-y)}}{q^2-M^2 +i\epsilon}
\left(\delta_{kl}-\frac{q_k q_l}{\vec{q}^2+M^2}\right)~.
\end{eqnarray}
 Together with the instantaneous  interaction (see the last term  in
 the reduced Lagrangian (\ref{redL})) this
 leads to the following current-current interaction
 \begin{equation}\label{mvecprop}
D^R_{\mu\nu}(q)J^\mu J^\nu = J_0 \frac{1}{\vec{q}^2+M^2} J_0 +J_i
J_j\left(\delta_{ij}-\frac{q_i q_j}{\vec{q}^2+M^2}\right)
\frac{1}{q^2-M^2}~.
\end{equation}
It is the generalization of the complete radiation  photon
propagator (\ref{wr}) in QED. The propagator (\ref{mvecprop})
contains the instantaneous part as a analogue of the Coulomb
instantaneous one in the propagator (\ref{wr}).

As it was shown in  \cite{hpp}, the Lorentz transformations of
classical radiation variables coincide with the  quantum ones  and
they both (quantum and classical) correspond to the transition to
another Lorentz reference frame distinguished by another
time-axis, where the relativistic covariant
propagator takes the
form
\begin{equation}\label{gv2}
D^{R}_{\mu\nu}(q|n)=\frac{n_{\mu}n_{\nu}}{M_W^2+|{q^{\perp}}^2|}
-\left(\delta^{\perp}_{\mu\nu}-\frac{q^{\perp}_{\mu}q^{\perp}_{\nu}}
{M_W^2+|{q^{\perp}}^2|}\right)\frac{1}{q^{2}-M_W^2+i\epsilon}
\end{equation}
with
 $
 q^{\perp}_{\mu}=q_{\mu}-n_{\mu}(qn),~~~
 \delta^{\perp}_{\mu\nu}=\delta_{\mu\nu}-n_{\mu}n_{\nu}
 $.
 Therefore, we shall use the radiation propagator (\ref{gv2}), where
 $n_{\mu}$ is treated as the initial datum operator   (\ref{3-al}),
instead of the conventional frame free massive vector
propagator \cite{db}
 \begin{equation}\label{mPhotprop}
 J^{\mu}D^F_{\mu\nu}(q)J^{\nu}=
-\frac{1}{q^2-M^2}J^{\mu}\left(g_{\mu\nu}-\frac{q_\mu q_\nu}{M^2}
\right)J^{\nu}~
\end{equation}
In contrast to this conventional frame free massive vector
propagator
the radiation propagator (\ref{mvecprop}) is regular in the limit
$M\rightarrow 0$ and is well behaved for large momenta.

For a better comparison with the conventional covariant propagator
(\ref{mPhotprop}) we rewrite the propagator (\ref{mvecprop}) in
the alternative form
\begin{equation}
D^R_{\mu\nu}(q)J^\mu J^\nu = -\frac{1}{q^2-M^2}\left(J_{\nu}^2
+\frac{(J_iq_i)^2-(J_0q_0)^2}{(\vec{q}^2+M^2)}\right)~.
\label{mvecprop2}
\end{equation}
Hence we see that for massive QED, where the vector field is
coupled to a conserved current $(q_{\mu}J^{\mu}=0)$, we find that
the effective current-current interactions mediated by the
propagator of the radiation  fields (\ref{mvecprop}) and by the
conventional frame free propagator (\ref{mPhotprop}) coincide
\begin{equation}
J^{\mu}D^R_{\mu\nu}J^{\nu}=J^{\mu}D^F_{\mu\nu}J^{\nu}~.
\end{equation}
 If the  current is not conserved $J_0q_0\not =J_kq_k$,
 the 
 radiation field variables
 with the propagator (\ref{mvecprop})
 (or  (\ref{gv2}))
 is inequivalent to   the frame free variables
 with the propagator  (\ref{mPhotprop}) because different
variables corresponds to different {\it gauges of physical sources}.

  In the next Section,
 we show  the dependence of
 physical results on {\it gauge of physical sources}.

\section{Observational tests}

\subsection{Chiral Bosonization of EW Interaction}

 In order to illustrate the inequivalence
 of  Dirac's method (\ref{redL})
  and the frame free one
 \cite{db},  we can consider the $K \to \pi$ transition amplitude
 and
 the kaon decays, where the dominance of the instantaneous
 interactions was recently predicted \cite{05}.
 Recall that the observation of kaon weak decays has been crucial
 for the modern theory of particle physics \cite{mikulec}.

 It was accepted to  describe  weak decays  in the framework of
 electroweak (EW) theory at the quark QCD level
 including  current vector boson weak interactions \cite{am,va}
 \be\label{ch111}
 \mathcal{L}_{(J)}=-(J^{-}_{\mu}W^{+}_{\mu}
 +J^{+}_{\mu}W^{-}_{\mu})=
 -\frac{e}{2\sqrt{2}\sin\theta_{W}}(\underline{J}^{-}_{\mu}W^{+}_{\mu}
 +\underline{J}^{+}_{\mu}W^{-}_{\mu}) ,
 \ee
 where
 $\underline{J}^{+}_{\mu}=\bar{d}'\gamma_{\mu}(1-\gamma_{5})u; \quad \bar{d}'=d
 \cos\theta_{C}+s\sin\theta_{C}, \theta_{C}$ is a Cabbibo angle
 $\sin\theta_{C}=0.223$.

 However, a consistent theory at large distances of QCD is not yet
 constructed up to now. Therefore, now  the most effective method
 of analysis  of kaon decay physics
 \cite{bvp,ecker,belkov01,CP-enhancement} is the chiral
 perturbation theory \cite{vp1,gs} (the list of arguments in
 favor of this perturbation theory is given in Appendix B).

 The quark content of $\pi^{+}$ and $K^{+}$
 mesons $\pi^{+}=(\bar{d},u), K^{+}=(\bar{s},u),
  \overline{K}^{0}=(\bar{s},d) $  leads to the
 effective chiral hadron currents $\underline{J}^{\pm}_\mu$ in
 the  Lagrangian (\ref{ch111})
 \be\label{chl1}
 \underline{J}^{\pm}_{\mu}=[\underline{J}^1_{\mu}{\pm}i\underline{J}^2_{\mu}]\cos\theta_{C}\,+
 [\underline{J}^4_{\mu}{\pm}i\underline{J}^5_{\mu}]\,\sin\theta_{C}\,,
 \ee
 where using the Gell-Mann matrixes $\lambda^k$ one can define
 the meson current as \cite{vp1}
 \be\label{c3k}
 i\sum\limits \lambda^k
 \underline{J}^k_{\mu}=i\lambda^k(V^{k}_{\mu}-A^{k}_{\mu})^{k}=F^2_\pi
 e^{i\xi}\partial_\mu e^{-i\xi},
 \ee
\be\label{c4}
 \xi=F_\pi^{-1}\sum\limits_{k=1}^{8}M^k\lambda^k=F_\pi^{-1}\left(%
\begin{array}{ccc}
  \pi^0+\dfrac{\eta}{\sqrt{3}} & \pi^+\sqrt{2} & K^+\sqrt{2} \\
  \pi^-\sqrt{2}  & -\pi^0+\dfrac{\eta}{\sqrt{3}} & K^0\sqrt{2} \\
  K^-\sqrt{2} & \overline{K}^0\sqrt{2} & -\dfrac{2\eta}{\sqrt{3}} \\
\end{array}%
\right).
 \ee
 In the first orders in mesons one can write

 \be \label{chl2}
 V^{-}_{\mu}=\sqrt{2}\,\,[\,\sin\theta_{C}\,
 (K^{-}\partial_{\mu}\pi^{0}-\pi^{0}\partial_{\mu}K^{-} )\,
 +\cos\theta_{C}\,(\pi^{-}\partial_{\mu}\pi^{0}-
 \pi^{0}\partial_{\mu}\pi^{-})\,]+...
  \ee
  and
  \be\label{chl3}
  A^{-}_{\mu}=\sqrt{2}\,F_{\pi}\,(
 \partial^{\mu}K^{-}\sin\theta_{C} +
 \partial^{\mu}\pi^-\cos\theta_{C})+...;
 \ee
 here $F_{\pi}\simeq  92$ MeV.
The right form of chiral Lagrangian of the electromagnetic
interaction of mesons  can be constructed by the covariant
derivative
 $
 \partial_{\mu}\chi^{\pm}\to D_{\mu}\chi^{\pm}
\equiv(\partial_{\mu}\pm ieA_{\mu})\chi^{\pm},
 $
 where $\chi^{\pm}=K^{\pm},\pi^{\pm}$.


We suppose also that the quark content of the
  mesons determines hadronization
 of QCD \cite{5,6}  conserving its
 chiral and gauge symmetries.



\subsection{$K \to \pi$ Transition Amplitude and the Rule $\Delta
T=\dfrac12$}

\begin{figure}
\centering
\begin{minipage}[c]{0.45\hsize}
\begin{picture}(200,100)(0,0)
\Vertex(70,50){5}\Vertex(130,50){5}
\ZigZag(70,50)(130,50){5}{5}\ArrowLine(130,50)(200,50)
\SetWidth{2.0} \ArrowLine(0,50)(70,50)
\Text(5,90)[]{$(a)$}\Text(35,30)[]{$K^{+}(k)$}
\Text(100,30)[]{$W^{+}(k)$}\Text(165,30)[]{$\pi^{+}(k)$}
\end{picture}
\end{minipage}\hspace*{5mm}
\begin{minipage}[c]{0.45\hsize}
\begin{picture}(200,100)(0,0)
\Vertex(70,50){5}\Vertex(130,50){5}\CArc(100,50)(30,0,180)
\ZigZag(70,50)(130,50){5}{5}\ArrowLine(130,50)(200,50)
\SetWidth{2.0} \ArrowLine(0,50)(70,50)
\Text(5,90)[]{$(b)$}\Text(35,30)[]{$K^{+}(k)$}\Text(95,90)[]{$\pi^{0}(k+l)$}
\Text(100,30)[]{$W^{+}(-l)$}\Text(165,30)[]{$\pi^{+}(k)$}
\end{picture}\end{minipage}
\caption{Axial (a) and vector (b) current contribution into
$K^+\to \pi^+$ transition} \label{1ac}\end{figure}
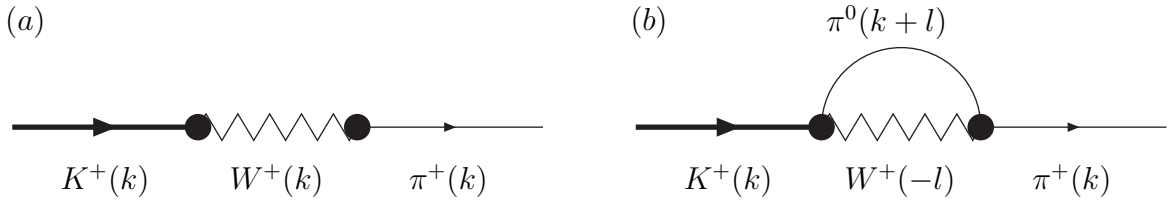

 Let us consider the $K^+\to \pi^+$ transition amplitude
\be\label{kp21}
 <\pi^+|-i\int dx^4dy^4
 J^\mu(x)D_{\mu\nu}^W(x-y)J^\nu(y)|K^+>
 =i(2\pi)^{4}\delta^{4}(k-p)G_{\rm EW}\Sigma(k^{2}),
 \ee
 in the first order of the EW perturbation theory in the Fermi
 coupling constant
 \be\label{g}
  G_{\rm EW}= \frac{\sin\theta_{C} \cos\theta_{C}}{8
  M^{2}_{W}}\frac{e^{2}}{\sin^{2}\theta_{W}}\equiv
  \sin\theta_{C} \cos\theta_{C}\frac{G_F}{\sqrt{2}},
 \ee
 comparing two different W-boson field propagators,
 the  radiation (R) propagator (\ref{gv2})
\begin{equation}\label{1gv2}
 D^W_{\mu\nu}(q)~\to~
 D^{R}_{\mu\nu}(q|n)=\frac{n_{\mu}n_{\nu}}{M_W^2+|{q^{\perp}}^2|}
-\left(\delta^{\perp}_{\mu\nu}-\frac{q^{\perp}_{\mu}q^{\perp}_{\nu}}
{M_W^2+|{q^{\perp}}^2|}\right)\frac{1}{q^{2}-M_W^2+i\epsilon},
\end{equation}
where
 $ q^{\perp}_{\mu}=q_{\mu}-n_{\mu}(qn),~
 \delta^{\perp}_{\mu\nu}=\delta_{\mu\nu}-n_{\mu}n_{\nu}$ in
the  frame $n_\mu = k_\mu/\sqrt{k^2_\mu}$,
and the frame free (F) propagator
(\ref{mPhotprop})
 \begin{equation}\label{1mPhotprop}
D^W_{\mu\nu}(q)~\to~~ D^F_{\mu\nu}(q)=
-\frac{1}{q^2-M_W^2}\left(g_{\mu\nu}-\frac{q_\mu q_\nu}{M_W^2}
\right)~.
\end{equation}
 These propagators give the expressions
 corresponding to the diagrams in Fig \ref{1ac}
\bea\label{kp2}
  \Sigma(k^{2})&\to&\Sigma^R(k^{2})=2F^2_{\pi}k_\mu
 D^{R}_{\mu\nu}(k|n)k_\nu +2i\int \frac{d^4q}{(2\pi)^4}
 \frac{(2k_\mu+q_\mu)D^{R}_{\mu\nu}(-q|n)(2k_\nu+q_\nu)}{(k+q)^2
 -m^2_\pi+i\epsilon}, \\
 \label{fkp2}
  \Sigma(k^{2})&\to&\Sigma^F(k^{2})=2F^2_{\pi}k_\mu
 D^{F}_{\mu\nu}(k)k_\nu +2i\int \frac{d^4q}{(2\pi)^4}
 \frac{(2k_\mu+q_\mu)D^{F}_{\mu\nu}(-q)(2k_\nu+q_\nu)}{(k+q)^2
 -m^2_\pi+i\epsilon}=\\\label{Fkp2}
&=& 2k_\mu^2\frac{F^2_{\pi}}{M_{W}^{2}}+
  \int \frac{d^4q}{(2\pi)^4}
\frac{(2k_\mu+q_\mu)(2k_\nu+q_\nu)}{(M_W^2-q^2-i\epsilon)[(k+q)^2
-m^2_\pi+i\epsilon]}\left[\delta^{\mu\nu}-\frac{q^\mu
q^\nu}{M_W^2}\right]~.
 \eea
 Both versions R and F coincide in the case of the axial contribution
 corresponding to the first diagram in Fig.  \ref{1ac}
 and both they reduce to
  the instantaneous interaction contribution, because
$$
k^\mu k^\nu D^F_{\mu\nu}(k)\equiv k^\mu k^\nu
D^R_{\mu\nu}(k|n)=\frac{k^2}{M^2_W}.
$$
 However, in the case of the vector contribution
 corresponding to the second diagram in Fig.  \ref{1ac}
 the radiation version differs from the frame free
 one (\ref{Fkp2})\footnote{The Faddeev equivalence theorem \cite{f1}
 is not valid, because
  the vector current $J_\mu = K\partial_\mu \pi-\pi\partial_\mu K$
  becomes the vertex $\Gamma_\mu = K\partial_\mu D_\pi-D_\pi\partial_\mu K$,
   where one of fields
  is replaced by its propagator $\Box D_\pi =\delta (x)$, and
  $\partial_\mu \Gamma^\mu \not = 0$.}.

  In contrast to the frame free version (\ref{Fkp2}),
  two radiation variable diagrams shown in Fig. \ref{1ac} in  the
  frame $n_\mu \sim k_\mu=(k_0,0,0,0)$
  are reduced to
  the instantaneous interaction contribution
 \be\label{kp1}
 i(2\pi)^{4}\delta^{4}(k-p)G_{\rm EW}\Sigma^R(k^{2})=
 <\pi^+|-i\int dx^4
 J_0(x)\frac{1}{\triangle-M_W^2}J_0(x)|K^+>
 \ee
  with the normal ordering of the pion fields, so that
 \be\label{0Rkp1}
 \Sigma^R(k^{2}) = 2k_\mu^2\left[\frac{F^2_{\pi}}{M_{W}^{2}}+
  \frac{1}{(2\pi)^3}\int\frac{d^3l}{2E_\pi(\vec{l})}
  \frac{1}{M^2_W+\vec{l}^2}\right];
 \ee
 here
 $E_{\pi}(\vec{l})=\sqrt{m^{2}_{\pi}+\vec{l}^{2}}$ is the
 energy of $\pi$-meson.
 The reduction of the radiation variable loop diagrams
  to the instantaneous interaction ones demonstrates that
  the radiation variables allow to separate the low-energy
  region from the high-energy one. This separation
    justifies
 the application of the low-energy chiral perturbation theory  \cite{fpp},
 where the summation of the chiral series can be
 considered here as
  the meson form factors
  \cite{bvp,belkov01,CP-enhancement}.
  Finally, the radiation variables  (\ref{kp2}) and (\ref{kp1})
 give the  result 
 \be\label{Rkp1}
 \Sigma^R(k^{2}) = 2k_\mu^2\left[\frac{F^2_{\pi}}{M_{W}^{2}}+
  \frac{1}{(2\pi)^3}\int\frac{d^3l}{2E_\pi(\vec{l})}
  \frac{f_{K\pi W}(-\vec{l}^2)f_{\pi\pi W}(-\vec{l}^2)}{M^2_W+\vec{l}^2}\right]
 \equiv
  2k_\mu^2\frac{F^2_{\pi}}{M_{W}^{2}}g_8;
 \ee
 here
  $f^{V}_{K\pi W}(-\vec{l^{2}})$ and
 $f^{V}_{\pi \pi W}(-\vec{l^{2}})$ are the weak vector form factors
 introduced  in  the chiral perturbation theory \cite{fpp}
as the natural regularization
 of the low-energy meson  instantaneous interactions by the next orders,
 and \be\label{vv3}
 g_8=1+\frac{M^{2}_{W}}{(2\pi)^{2}F^2_{\pi}}\int\frac{d|\vec
l| ~~ \vec{l^{2}}}{E_{\pi}(\vec{l})}\frac{f^{V}_{K\pi
W}(\vec{-l^{2}})f^{V}_{\pi
 \pi W}(-\vec{l}^{2})}{M^{2}_{W}+\vec{l}^{2}},
 \ee
 is
 the parameter of   the enhancement of the probability
 of the axial $K^+ \to \pi^+$ transition. The relation
 (\ref{vv3}),
 where $g_8=5.1$ \cite{ecker,7a}, can be considered as the low-energy sum rule
 for meson form factors \cite{duby} in the space-like region.



 This result shows that the instantaneous vector interaction
 can increase the axial interaction $K^+\to \pi^+$ transition
 in $g_8$ times and give a new term describing the
  $K^0\to \pi^0$ transition proportional
 to $g_8-1$.

 Using
  the covariant perturbation theory \cite{pv}\ developed as the
 series $\underline{J}_\mu^k(\gamma
 \oplus\xi)=\underline{J}_\mu^k(\xi)+ F^2_\pi
 \partial_\mu \gamma^k +\gamma^if_{ijk}\underline{J}_\mu^j(\xi)+O(\gamma^2)$
 with respect to quantum fields $\gamma$ added to $\xi$
 as  the product $e^{i\gamma}e^{i\xi}\equiv e^{i(\gamma \oplus\xi)}$
  \cite{pv}, one can see that the normal ordering $N(\vec x-\vec y)=<0|\gamma^i(x)\gamma^{i'}(y)|0>=
  \delta^{ii'}\int d^3l e^{i\vec l\cdot (\vec x -\vec y)}/(2E_\pi(\vec l))$
  in the product of the currents can lead to an effective Lagrangian
$$
 \int d^3z
 \frac{N(\vec z)e^{-M_W|\vec z|}}{4\pi|\vec z|}
 \underline{J}_\mu^j(\xi(x))\underline{J}_\mu^{j'}(\xi(z+x))
 [(f_{ij1}+i f_{ij2})(f_{ij'4}-if_{ij'5})+h.c]<0|\gamma^i(x)\gamma^{i'}(y)|0>.
 $$
 In the  limit $M_W\to \infty$, one can neglect the
  dependence of the current on $\vec z$, and we get
  the effective  Lagrangians  \cite{kp}
 \be \mathcal{L}_{(\Delta T=\frac{1}{2})}=
\frac{G_{F}}{\sqrt{2}}g_{8}\cos\theta_{C}\sin\theta_{C}
\Big[(\underline{J}^1_{\mu}+i\underline{J}^2_{\mu})
(\underline{J}^4_{\mu}-i\underline{J}^5_{\mu})-
(\underline{J}^3_{\mu}+\frac{1}{\sqrt{3}}\underline{J}^8_{\mu})
(\underline{J}^6_{\mu}-i\underline{J}^7_{\mu})+h.c.\Big], \ee \be
\mathcal{L}_{(\Delta T=\frac{3}{2})}=
\frac{G_{F}}{\sqrt{2}}\cos\theta_{C}\sin\theta_{C}
\Big[(\underline{J}^3_{\mu}+\frac{1}{\sqrt{3}}\underline{J}^8_{\mu})
(\underline{J}^6_{\mu}-i\underline{J}^7_{\mu})+h.c.\Big]. \ee
 This Lagrangian at the level of the tree diagrams
 describes the nonleptonic decays in
satisfactory agreement with experimental data within the
 accuracy  $20\div 30\%$  \cite{vp1,kp}.

\subsection{The  $K^+\to\pi^++l+\bar l$ amplitude}

\begin{figure}
\centering
\begin{minipage}[c]{0.45\hsize}
\begin{picture}(200,100)(0,0)\Vertex(150,50){5}\Vertex(50,50){5}\Vertex(100,50){5}
\ZigZag(50,50)(100,50){5}{5}\ArrowLine(100,50)(150,50)\ArrowLine(150,50)(200,50)
\Photon(150,50)(150,100){5}{3} \SetWidth{2.0}
\ArrowLine(0,50)(50,50)
\Text(5,90)[]{$(a)$}\Text(25,30)[]{$K^{+}(k)$}
\Text(125,30)[]{$\pi^{+}(k)$}\Text(175,30)[]{$\pi^{+}(p)$}\Text(75,30)[]{$W^{+}(k)$}
\Text(170,85)[]{$\gamma^{*}(q)$}
\end{picture}
\end{minipage}
\hspace*{5mm}
\begin{minipage}[c]{0.45\hsize}
\begin{picture}(200,100)(0,0)\Vertex(150,50){5}\Vertex(50,50){5}\Vertex(100,50){5}
\ZigZag(100,50)(150,50){5}{5}\ArrowLine(150,50)(200,50)
\Photon(50,50)(50,100){5}{3} \SetWidth{2.0}
\ArrowLine(0,50)(50,50)\ArrowLine(50,50)(100,50)\Text(5,90)[]{$(b)$}
\Text(25,30)[]{$K^{+}(k)$}
\Text(125,30)[]{$W^{+}(p)$}\Text(175,30)[]{$\pi^{+}(p)$}\Text(75,30)[]{$K^{+}(p)$}
\Text(70,85)[]{$\gamma^{*}(q)$}
\end{picture}
\end{minipage}
\begin{minipage}[c]{0.45\hsize}
\begin{picture}(200,100)(0,0)
\Vertex(70,50){5}\Vertex(130,50){5}
\ZigZag(70,50)(130,50){5}{5}\ArrowLine(130,50)(200,50)
\Photon(130,50)(130,100){5}{3} \SetWidth{2.0}
\ArrowLine(0,50)(70,50)
\Text(5,90)[]{$(c)$}\Text(35,30)[]{$K^{+}(k)$}
\Text(100,30)[]{$W^{+}(k)$}\Text(165,30)[]{$\pi^{+}(p)$}
\Text(150,85)[]{$\gamma^{*}(q)$}
\end{picture}
\end{minipage}\hspace*{5mm}
\begin{minipage}[c]{0.45\hsize}
\begin{picture}(200,100)(0,0)
\Vertex(70,50){5}\Vertex(130,50){5}
\ZigZag(70,50)(130,50){5}{5}\ArrowLine(130,50)(200,50)
\Photon(70,50)(70,100){5}{3} \SetWidth{2.0}
\ArrowLine(0,50)(70,50)
\Text(5,90)[]{$(d)$}\Text(35,30)[]{$K^{+}(k)$}
\Text(100,30)[]{$W^{+}(p)$}\Text(165,30)[]{$\pi^{+}(p)$}
\Text(90,85)[]{$\gamma^{*}(q)$}
\end{picture}
\end{minipage}
\caption{Axial current contribution} \label{ac}\end{figure}
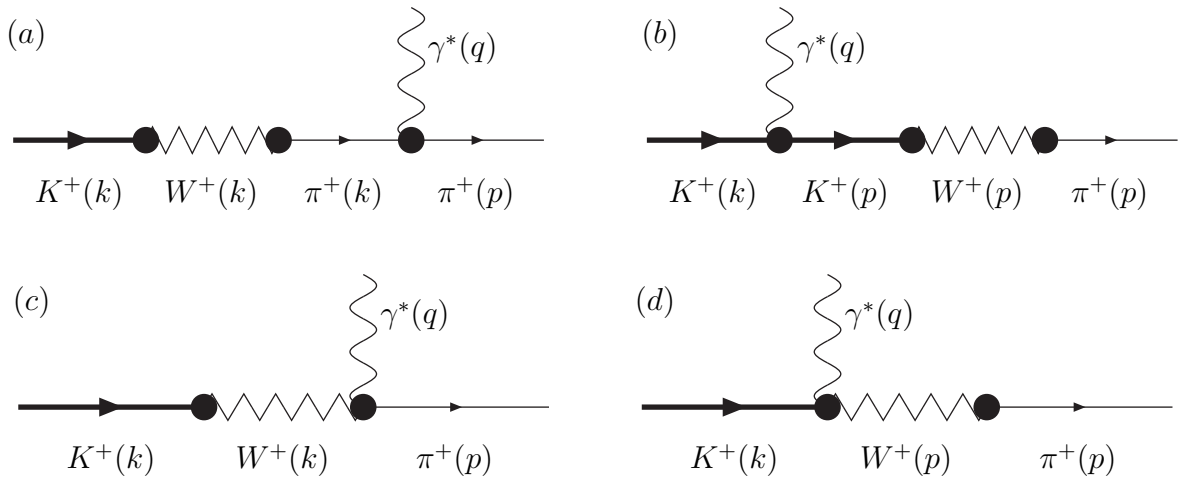

 The result of calculation of the axial contributions to the
amplitude of the process $K^+\to\pi^++l+\bar l$ within the
framework of chiral Lagrangian (\ref{ch111}), (\ref{chl1})
including phenomenological meson form factors shown  in Fig.
 \ref{ac} as a fat dot  takes the form

 \be\label{ampl}
  T^{AA}_{(K^+\to \pi+l^+l^-)}= 2eG_{\rm EW}L_{\nu}
 D^{\gamma(rad)}_{\mu\nu}(q)(k_{\mu }+p_{\mu })\,\,\,
 t^{AA}(q^2,k^2,p^2),
 \ee
 where $G_{\rm EW}$ is the constant (\ref{g}),
  $L_{\mu}=\bar{l}\gamma_{\mu}l$ is leptonic current,
 \be\label{tp}
 t^{AA}(q^2,k^2,p^2)=F^2_{\pi}
 \left[\frac{f^{V}_{\pi}(q^{2})k^{2}}{m^{2}_{\pi}-k^{2}-
 i\epsilon}+\frac{f^{V}_{K}(q^{2})p^{2}}{M^{2}_{K}-p^{2}-i\epsilon}+
 \frac{f^{A}_{K}(q^{2})+f^{A}_{\pi}(q^{2})}{2}\right],
 \ee
 and
 $f^{(A,V)}_{\pi,K}(q^{2})$ are meson form factors.
 In this expression the propagator
 $D_{\mu\nu}^{{R}}(p|n)$ given by Eq.
 (\ref{gv2}) keeps only the covariant instantaneous
 part
  $\dfrac{p_\mu p_\nu}{p^2 M^2_W}$.

 On the mass-shell the sum (\ref{tp})  takes the form
 \bea\label{t0}
 t^{AA}(q^2,M_K^2,m_\pi^2)&=&\\\nonumber
 = t(q^2)&=&F^2_\pi  \Bigg[\frac{f^{A}_{K}(q^{2})+f^{A}_{\pi}(q^{2})}{2}
 - f^{V}_{\pi}(q^{2})+
 [f^{V}_{K}(q^{2})- f^{V}_{\pi}(q^{2})]\frac{m_\pi^2}{M_K^2-m_\pi^2}
  \Bigg]~.
 \eea

As can be seen in \cite{bvp,ecker}, the amplitude for $K^{+}\to
\pi^{+}+l+\bar l $ vanishes at the tree level, where  form factors
are equal to unit
  $
t(q^2)|_{f^{V}=f^{A}=1}=0.
 $ 
 It is well known that these form factors in the limit $q^2\to 0$
  take the form
  $
 f^{A,V}_{\pi,K}(q^{2})=1-\frac{q^{2}}{6}<r^{2}>^{(A,V)}_{\pi,K}
 $, 
  where  $<r^{2}>^{(A,V)}_{\pi,K}$ are the axial and vector
 mean square radii of mesons, respectively, in agreement with
 the chiral perturbation theory \cite{bvp,vp1,fpp}.

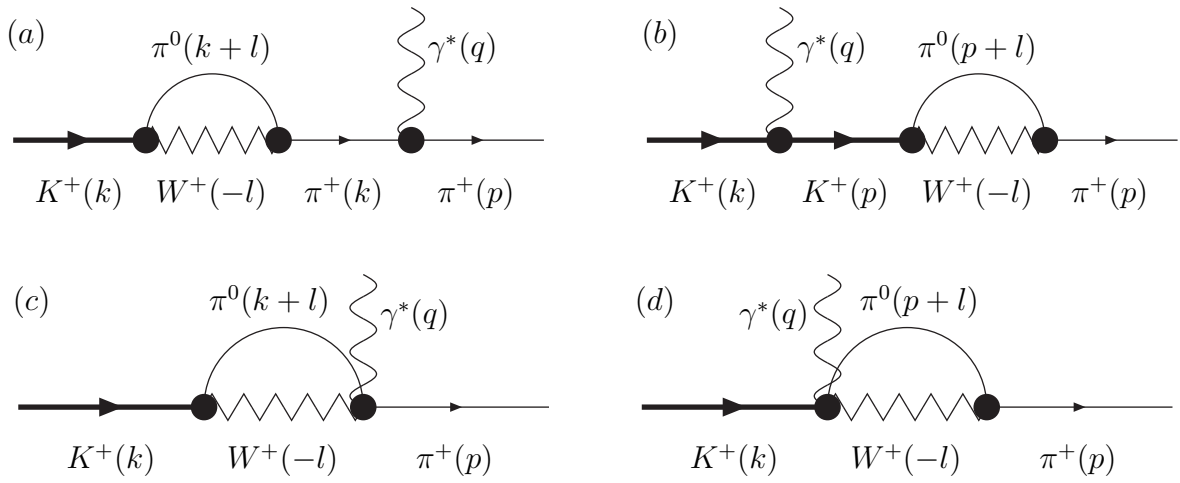
\begin{figure}
\centering
\begin{minipage}[c]{0.45\hsize}
\begin{picture}(200,100)(0,0)\Vertex(150,50){5}\Vertex(50,50){5}\Vertex(100,50){5}\CArc(75,50)(25,0,180)
\ZigZag(50,50)(100,50){5}{5}\ArrowLine(100,50)(150,50)\ArrowLine(150,50)(200,50)
\Photon(150,50)(150,100){5}{3} \SetWidth{2.0}
\ArrowLine(0,50)(50,50) \Text(5,90)[]{$(a)$}
\Text(25,30)[]{$K^{+}(k)$}
\Text(125,30)[]{$\pi^{+}(k)$}\Text(75,85)[]{$\pi^{0}(k+l)$}\Text(175,30)[]{$\pi^{+}(p)$}\Text(75,30)[]{$W^{+}(-l)$}
\Text(170,85)[]{$\gamma^{*}(q)$}
\end{picture}
\end{minipage}
\hspace*{5mm}
\begin{minipage}[c]{0.45\hsize}
\begin{picture}(200,100)(0,0)\Vertex(150,50){5}\Vertex(50,50){5}\Vertex(100,50){5}\CArc(125,50)(25,0,180)
\ZigZag(100,50)(150,50){5}{5}\ArrowLine(150,50)(200,50)
\Photon(50,50)(50,100){5}{3} \SetWidth{2.0}
\ArrowLine(0,50)(50,50)\ArrowLine(50,50)(100,50)\Text(5,90)[]{$(b)$}
\Text(25,30)[]{$K^{+}(k)$}\Text(125,85)[]{$\pi^{0}(p+l)$}
\Text(125,30)[]{$W^{+}(-l)$}\Text(175,30)[]{$\pi^{+}(p)$}\Text(75,30)[]{$K^{+}(p)$}
\Text(70,85)[]{$\gamma^{*}(q)$}
\end{picture}
\end{minipage}
\begin{minipage}[c]{0.45\hsize}
\begin{picture}(200,100)(0,0)
\Vertex(70,50){5}\Vertex(130,50){5}\CArc(100,50)(30,0,180)
\ZigZag(70,50)(130,50){5}{5}\ArrowLine(130,50)(200,50)
\Photon(130,50)(130,100){5}{3} \SetWidth{2.0}
\ArrowLine(0,50)(70,50)
\Text(5,90)[]{$(c)$}\Text(35,30)[]{$K^{+}(k)$}\Text(95,90)[]{$\pi^{0}(k+l)$}
\Text(100,30)[]{$W^{+}(-l)$}\Text(165,30)[]{$\pi^{+}(p)$}
\Text(150,85)[]{$\gamma^{*}(q)$}
\end{picture}
\end{minipage}\hspace*{5mm}
\begin{minipage}[c]{0.45\hsize}
\begin{picture}(200,100)(0,0)
\Vertex(70,50){5}\Vertex(130,50){5}\CArc(100,50)(30,0,180)
\ZigZag(70,50)(130,50){5}{5}\ArrowLine(130,50)(200,50)
\Photon(70,50)(70,100){5}{3} \SetWidth{2.0}
\ArrowLine(0,50)(70,50)\Text(5,90)[]{$(d)$}
\Text(35,30)[]{$K^{+}(k)$}\Text(105,90)[]{$\pi^{0}(p+l)$}
\Text(100,30)[]{$W^{+}(-l)$}\Text(165,30)[]{$\pi^{+}(p)$}
\Text(50,85)[]{$\gamma^{*}(q)$}
\end{picture}
\end{minipage}
\caption{Vector current contribution} \label{vc}\end{figure}

 The result of calculation of vector contributions to the amplitude
 of the process $K^+\to\pi^++l+\bar l$ within
 the framework of chiral Lagrangian (\ref{ch111}), (\ref{c3})
 including phenomenological meson form factors  shown  in Fig.
 \ref{vc} as a fat dot  takes the form
 \be T^{VV{(\rm rad)}}_{(K^+\to \pi+l^+l^-)}=
2eG_{\rm EW}L_{\nu} D^{\gamma(rad)}_{\mu\nu}(q)(k_{\mu } +p_{\mu
})\,\,\, t^{VV}(q^2,k^2,p^2), \label{ampl1}
 \ee
 where we use the propagator (\ref{gv2}) in the initial datum
frame (\ref{3-al})

 \begin{eqnarray}
 \nonumber
  t^{VV}(q^2,k^2,p^2)&=&\frac{1}{2(2\pi)^{2}}\int\frac{d|\vec l| ~~
\vec{l^{2}}}{E_{\pi}(\vec{l})} \Bigg\{{f^{V}_{K\pi
W}(\vec{l^{2}})f^{V}_{\pi
 \pi W}(\vec{l}^{2})}\Bigg[
\frac{f^V_{\pi}(q^2)k^{2}} {m^{2}_{\pi}-k^{2}-i\epsilon} +
 \\\nonumber
 +\frac{f^V_{K}(q^2)p^{2}}{M^{2}_{K}-p^{2}-i\epsilon}\Bigg]&+&
   \frac{f_\pi^A(q^2)f^V_{K \pi W}(\vec{l^{2}})
   f_{\pi \pi W\gamma}^{V}(\vec{l^{2}})+
   f_K^A(q^2)f^V_{\pi \pi W}(\vec{l^{2}})
   f_{K \pi W\gamma}^{V}(\vec{l^{2}})}{2}
   \Bigg\};
\end{eqnarray}
 here $f_{\pi \pi W\gamma}^{V}(\vec{l^{2}}),~f_{K \pi
W\gamma}^{V}(\vec{l^{2}})$ are the four particle interaction form
factors which should coincide with the three particle interaction
 form factors $f_{\pi \pi W\gamma}^{V}=f_{\pi \pi W}^{V},~
 f_{K \pi W\gamma}^{V}=f_{K \pi W}^{V}$ in accordance with
 the gauge invariance of the strong interactions.

 Using the result (\ref{vv3}) one can write
 on the mass shell $k^{2}=M_{K}^{2}, p^{2}=m_{\pi}^{2}$:
\bea\label{tvv}
t^{VV}(q^{2},M_{K}^{2},m_{\pi}^{2})&=&(g_8-1)t(q^{2}),\\
t^{AA}(q^{2},M_{K}^{2},m_{\pi}^{2})+
t^{VV}(q^{2},M_{K}^{2},m_{\pi}^{2})&=&g_8\,\,t(q^{2}),
 \eea
 where $g_8,t(q^{2})$ are given by Eqs. (\ref{vv3}) and (\ref{t0}), respectively.

 Finally, the result of calculation of axial and vector contributions to
 the amplitude of the process $K^+\to\pi^++l+\bar l$ within the
 framework of chiral Lagrangian (\ref{chl1}),(\ref{chl2}) including
 phenomenological meson form factors takes the form
 \be\label{amplv}
  T^{\rm (rad)}_{(K^+\to \pi^+l^+l^-)}=2e g_8
  G_{\rm EW}   L_{\nu}
 D^{\gamma (rad)}_{\mu\nu}(q)(k_{\mu }+p_{\mu })\,\,t(q^2).
 \ee

 This amplitude leads to  the total decay rate  for the transition $K^+ \to
 \pi^+e^{+}e^{-}$
 \be\label{f2}
 \Gamma_{(K^+ \to \pi^+e^{+}e^{-})}((M_K-m_\pi)^2)=\overline{\Gamma}_{e^+e^-}
 \int\limits_{4m^2_e}^{(M_K-m_\pi)^2}
 {\frac{d q^2}{M_K^2} \rho(q^2)}F(q^2),
 \ee
 where
 \bea\label{gf3}
 \overline{\Gamma}_{e^+e^-}=\frac{(s_1c_1c_3)^2g_8^2G^2_F}{(4\pi)^4}
 \frac{\alpha^2 M_K^5}{24\pi}|_{g_8=5.1}=1.37 \times 10^{-22}
 \mbox{\rm GeV},
 \eea
 \bea\label{f3}\nonumber
 F(q^2)&=& \left[\frac{(4\pi)^2t(q^2)}{q^2}\right]^2=\\\label{1f3}
  &=&\left[\frac{(4\pi F_\pi)^2}{q^2}\right]^2
  \Bigg[\frac{f^{A}_{K}(q^{2})+f^{A}_{\pi}(q^{2})}{2}
 - f^{V}_{\pi}(q^{2})+
 [f^{V}_{K}(q^{2})-
 f^{V}_{\pi}(q^{2})]\frac{m_\pi^2}{M_K^2-m_\pi^2},
  \Bigg]^2
 \eea
and
 $$
 \rho(q^2)=\left(1-\frac{4m_l^2}{q^2}\right)^{1/2}\times\left(1+\frac{2m_l^2}{q^2}\right)~
 \lambda^{3/2}
 (1,q^2/M^2_K,m^2_\pi/M^2_K);
 $$
 here $\lambda (a,b,c)=a^2+b^2+c^2-2(ab+bc+ca)$,
  $s_1 c_1 c_3$ is the product of
  Cabibbo-Kobayashi-Maskawa matrix elements $V_{ud}V_{us}$.

 The
 mechanism of enhancement $|\Delta T| = \frac{1}{2}$
 considered in this paper can be generalized to a description
 of other processes including $K^+\to \pi^+ \nu \bar{\nu}$
 by replacing the $\gamma$-propagator by the Z- boson one, so that
 we have the relations
 \bea\label{dr}\nonumber
 &&\left[\frac{q^2}{(4\pi F_\pi)^2}\right]^2\!\!\!\frac{M^2_K}{\rho(q^2)
\overline{\Gamma}_{e^+e^-}}
\frac{d
 \Gamma_{(K^+ \to \pi^+e^{+}e^{-})}(q^2)}{
  \,\,\,\,d q^2}
=\\\nonumber
  &&=\left[\frac{M_Z^2}{(4\pi F_\pi)^2}\right]^2\frac{M^2_K}{\rho(q^2)
\overline{\Gamma}_{\nu\bar \nu}}
\frac{d
 \Gamma_{(K^+ \to \pi^+\nu \bar \nu)}(q^2)}{
  \,\,\,\,d q^2}=
\\\label{dr1}
 &&=
\Bigg[\frac{f^{A}_{K}(q^{2})\!+\!f^{A}_{\pi}(q^{2})}{2}
 - f^{V}_{\pi}(q^{2})+
 [f^{V}_{K}(q^{2})- f^{V}_{\pi}(q^{2})]\frac{m_\pi^2}{M_K^2-m_\pi^2}
  \Bigg]^2.
 \eea
Thus, exploiting
 the instantaneous weak interaction
 mechanism of enhancement in the
 $K \to \pi$ transition probability
 and QCD symmetry we derive the sum rule of EW vector
 meson form factors  given by Eqs. (\ref{vv3}), (\ref{t0}), (\ref{f2})
  and their relation to the differential $K \to \pi e^+ e^-$ decay
  rate (\ref{dr}).

\subsection{The form factor probe of
the differential $K \to \pi e^+ e^-$ decay
  rate}

  The estimation of
 the  meson loop contribution was made
 in \cite{ecker} where
  a  function  $\hat \phi^2(q^2)$ was used
   instead of the
   form factor rate  $F(q^2)$ (\ref{1f3})
  \bea\nonumber
   F(q^2)~~\Longrightarrow~~ \hat \phi^2(q^2)
  \eea
  It was shown
  that the values $g_8=5.1$, $\hat \phi(0)=1$  gave
the total decay rate $\Gamma_{(K^+ \to \pi^+e^{+}e^{-})}=1.91\times
 10^{-23}$ GeV
  in the satisfactory agreement with the experimental data
 $\Gamma_{(K^+ \to \pi^+e^{+}e^{-})}=1.44\pm 0.27\times
 10^{-23}$ GeV \cite{7a,b}. However, the main contribution
 goes from the baryon loops  \cite{bvp,belkov01,CP-enhancement}.
 Therefore, we discuss here
 the value of the differential $K \to \pi e^+ e^-$ decay
  rate (\ref{1f3}) in the chiral perturbation theory \cite{vp1,fpp} where
 both the pion loop contribution $ \Pi_\pi$ and baryon ones
 lead to the meson form factors and resonances \cite{bvp,fpp}
 in the Pad\'e-type approximation
 \bea\label{1vform}
 f_{\pi}^{V}(q^2)&=&1+M^{-2}_\rho q^2+
 \alpha_{0}\Pi_\pi(q^2)+...\simeq\frac{1}{1-M^{-2}_\rho q^2-
 \alpha_{0}\Pi_\pi(q^2)}
 \\\label{2vform}
 f_{\pi}^{A}(q^2)&\simeq&f_{K}^{A}(q^2)=1+M^{-2}_a
 q^2+...\simeq\frac{1}{1-M^{-2}_a q^2}
 \eea
where
 \be\label{m1}\alpha_0=\dfrac{m_\pi^2}{(4\pi F_\pi)^2}\dfrac{4}{3}\simeq
 \dfrac{2}{103},
 \ee
\bea\label{M1}
 \Pi_\pi(t)= (1-\bar t)\left(\dfrac{1}{\bar t}-1\right)^{1/2}
 \arctan\left(\dfrac{\bar t^{1/2}}{(1-\bar t)^{1/2}}\right)-1;
 ~~~~~0<t<(2m_\pi)^2
 \\
 \Pi_\pi(t)= \dfrac{\bar t-1}{2}\left(1-\frac{1}{\bar t}\right)^{1/2}
\left\{i\pi
 -\log
 \dfrac{\bar t^{1/2}+(\bar t-1)^{1/2}}{\bar t^{1/2}-(\bar t-1)^{1/2}}
 \right\}+1;
 ~~~~~(2m_\pi)^2<t
 \eea
 is the pion loop contribution \cite{vp1,fpp}, and
 $M_\rho=771$ MeV and
  $M_a=980$ MeV
 are  the values of resonance masses \cite{7a} that can be
 determined by the baryon loops  \cite{bvp,belkov01,CP-enhancement,fpp}.

 The constant approximation of the type of \cite{ecker}
 \be\label{M2}
 F(q^2=0)=F(0) =
  {(4\pi F_\pi)^4}\frac{[f_\pi^A(q^2)-f_\pi^V(q^2)]^2}{(q^2)^2}|_{q^2=0}=
  (4\pi F_\pi)^4[M_\rho^{-2}-M_a^{-2}]^2= 0.74
  \ee
  corresponds  to the value of $\Gamma_{(K^+ \to \pi^+e^{+}e^{-})}=1.41\times
 10^{-23}$ GeV
 in satisfactory agreement with
 the experimental data $\Gamma_{(K^+ \to \pi^+e^{+}e^{-})}=1.44\pm 0.27\times
 10^{-23}$ GeV \cite{7a,b}.

 The form factors (\ref{1vform}) and (\ref{2vform}) below
 the two particle threshold $q^2<4m^2_\pi$ determines
  the differential rate $F(q^2)$
 as a function of $q^2$
 \be\label{f4}
 F(q^2)=\frac{F(0)}{[1-M_\rho^{-2}q^2]^2[1-M_a^{-2}q^2]^2}~~\longrightarrow~~
 \frac{d}{d q^2}\log F(q^2)|_{q^2=0}\simeq 2[M_\rho^{-2}+M_a^{-2}]\simeq 5
 \mbox{\rm GeV}^{-2}
 \ee
 At the region above
 the two particle threshold $4m^2_\pi<q^2<(M_K-m_\pi)^2$ there is
  a jump of the differential rate $F(q^2)$ at the level of
 $5\div10$ \%
\be\label{f5}
 F(q^2)=(4\pi F_\pi)^4\dfrac{[M_\rho^{-2}-M_a^{-2}]^2+
 {\alpha^2_0\pi^2(q^2-4m_\pi^2)}/{(4q^6)}}
 {[1-M_\rho^{-2}q^2]^2[1-M_a^{-2}q^2]^2}
 \ee

 These results (\ref{M2}), (\ref{f4}), and (\ref{f5}) are  arguments that the detailed
  investigation of the differential $K \to \pi e^+ e^-$ decay
  rate in the NA48/2 CERN experiment
 can give us information about  the meson form factors.
 To make much more
 realistic analysis,
 we have to use the unitary and analytic
 model for meson form factors \cite{duby} describing very well all
 experimental data.


\section{Conclusion}

 The instantaneous interactions
 are inevitable consequence of the Dirac approach to gauge theories.
 Recall that Dirac
   eliminated all  fields with zero momenta (and
 possible negative contributions into the energy of the
 system) by resolution of the Gauss constraint
 and  its diagonalization,
 in order to  have
  a physical vacuum as a state with minimal
 energy in a comoving {\it inertial frame}.

   This elimination unambiguously  leads to the {\it radiation variables}
 associated  in QED with the Coulomb {\it gauge constraint}.
    However, in SM, the {\it radiation variables}
   of the massive vector bosons do not correspond to any whatsoever
   {\it gauge} constraint. The transition from
 the {\it radiation variables} with the instantaneous interactions
 to a frame free formulation \cite{db} without
 the instantaneous interactions is fulfilled by
 {\it gauge of physical sources}.
The change of the {\it gauge of physical sources}
 is just the
 point where the accepted frame free methods \cite{db}
 lose all instantaneous  interactions together with their
 physical effects, including the instantaneous bound states.
  The strong dependence of
  physical results on the {\it gauge of physical sources}
  does not
  mean the violation of the gauge-invariance principle.
 This means that  {\it frame free}  method has a region of validity
  restricted by the {\it initial datum free} processes of the
  type of the elementary particles scattering ones \cite{f1}, as
 it follows from  the physical meaning of the concept
 of the {\it frame} revealed  by its  full name
 {\it frame of reference to initial data}.

 In order to keep relativistic invariance
 of S-matrix elements between the instantaneous bound states,
   the instantaneous
 interactions are determined in an arbitrary frame, where its time-axis
is treated as the  operator of {\it initial data}
 acting
 in the complete set of the all asymptotic physical states
 including the bound states considered as irreducible representations of
 the Poincar\'e group  \cite{wigner}.

 In the paper, we compare this Dirac operator approach to SM
 with the acceptable frame free method \cite{db}
 using as example the $K^+ \to \pi^+$ transition
 and the semileptonic and  nonleptonic kaon decay
 probability amplitudes. This comparison shows us that the
Dirac approach to SM separates the low-energy contributions from
the high-energy ones and expresses
   these amplitudes  in
terms of the hadron electroweak form factors.
Therefore, the obtained  amplitudes allow  to
extract information about form factors of $\pi $ and $K$ mesons
from the $K^+ \to \pi^+$ transition in the $K^+ \to
\pi^+e^{+}e^{-}(\mu^{+}\mu^{-},\nu\bar{\nu})$ and $K^+ \pi^-\to
e^{+}e^{-}(\mu^{+}\mu^{-},\nu\bar{\nu})$ processes.
 The parameters of  the differential $K \to \pi e^+ e^-$ decay
  rate in the NA48/2 CERN experiment were  estimated in the chiral perturbation
theory.

  These results show  that the Dirac
   formulation
  of Standard Model can reveal new physical
  effects in  comparison with
  the frame free  formulation \cite{db} used now for
  describing observational data. Therefore, the problem of
  the Dirac formulation of SM becomes topical in the light of
  future precision experiments.

\section*{Acknowledgments}

The authors are grateful to
   B.M. Barbashov, D.Yu. Bardin, A.Di Giacomo, S.B. Gerasimov,
     G.V. Efimov,
    A.V. ~Efremov,  V.D. Kekelidze, E.A. Kuraev,
   V.B.~Priezzhev,    A.N. Sissakian, S.I.~Vinitsky
 and M.K. Volkov for fruitful discussions.
 The work was in part supported by the
 Slovak Grant Agency for Sciences VEGA, Gr.No.2/4099/26.

{\small
\section*{Appendix A: S-matrix for bound states in electrodynamics}

\renewcommand{\theequation}{A.\arabic{equation}}

\setcounter{equation}{0}


{\bf A.1. Spectral problem}

\vspace{0.1cm}

 Let us keep in the QED action only
 the instantaneous
 interaction  (\ref{15-2a})
 neglecting radiation corrections
  \begin{eqnarray} \label{act2}
 \!S^{(\rm rad)}_{\rm QED} [ \psi , \bar{\psi} ]\!
 = \!\int d^{4}x [ \bar{\psi}(x) ( i \rlap/{\partial} - m ) \psi(x) +
  \frac{1}{2} \int d^{4}y ( \psi(y) \bar{\psi}(x) ) {\cal
 K} ( x,y\mid n ) ( \psi(x) \bar{\psi}(y) ) ] ;
 \end{eqnarray}
 here
  \begin{eqnarray} \label{3-7}
 {\cal K}(x,y \mid n ) = \rlap/n^{\rm cf} V(z^{\perp})
  \delta(z \cdot n^{\rm cf} ) \rlap/n^{\rm cf},
  ( \rlap/n = n^{\mu}
 \gamma_{\mu}=\gamma \cdot n)
 \end{eqnarray}
is the instantaneous kernel depending on
   the relative coordinates $z=x-y$
 and $V(z^{\perp})=-\dfrac { e^2}{4\pi|z^{\perp}|}$
  depends on the
 transverse  components $z_{\mu}^{\perp} = z_{\mu} -
 n^{\rm cf}_{\mu} ( z \cdot n^{\rm cf} )$
 of the relative coordinate
 with respect to the time axis
  $
 n^{\rm cf}_\mu=(1,0,0,0)
 $.

 It seems that a most straightforward way for constructing S-matrix of
 bound states in gauge theories
 is the redefinition of action (\ref{act2}) in terms of bilocal
 fields by means of the Legendre transformation \cite{pre} of the
 current -- current interaction with a kernel ${\cal K}(x,y|n)$
 \begin{eqnarray}  \label{3-8}
 &&\frac{1}{2} \int d^{4}x d^{4}y  ( \psi(y) \bar{\psi}(x) )
 {\cal K}(x,y|n) ( \psi(x) \bar{\psi}(y) )  =  \\  =
  -&&\frac{1}{2}
 \int d^{4}x d^{4}y  {\cal M}(x,y) {\cal K}^{-1}(x,y|n) {\cal
 M}(x,y)   +  \int d^{4}x d^{4}y ( \psi(x) \bar{\psi}(y) )
 {\cal M}(x,y). \nonumber
 \end{eqnarray}
  and the Gauss functional integral over the bilocal fields
  in the generating functional 
 \bea\label{16-12}
 \!\!\!Z[\eta^{(\rm rad)},\!\bar\eta^{(\rm rad)}]
 \!\!\!=\!\!\!
 \int D{\cal M} D\psi D\bar{\psi}\exp\!\!
 \left\{\!(\! \psi \bar{\psi}, ( - G_{0}^{-1} + {\cal M})
 )\! -\!\!\! \frac{1}{2} ( {\cal M}, {\cal K}^{-1} {\cal M} )
 +i(\bar\psi\eta^{(\rm rad)})\!\!+\!\!(\bar\eta^{(\rm rad)}\psi)\!\!\right\};
 \eea
 here we used  the short - hand notation
 \begin{eqnarray} \label{short}
 \int d^{4}x d^{4}y \psi (y) \bar{\psi}(x) ( i
 \rlap/\partial - m^{0} ) \delta^{(4)} (x-y)
 &=&  ( \psi \bar{\psi} , - G_{0}^{-1} ) \, \, ,  \\
 \int d^{4}x d^{4}y ( \psi(x) \bar{\psi}(y) ) {\cal M}(x,y) &=& (
 \psi \bar{\psi}, {\cal M} ) .
 \end{eqnarray}
 The integration over
  the fermi fields $\psi,\bar\psi$ in (\ref{16-12}) leads to
  the generating functional
 \bea\label{16-13}
 Z_{FP}[\eta^{(\rm rad)},\bar\eta^{(\rm rad)}]=
 \int D{\cal M}\exp\left\{iW_{eff}[{\cal M}]
 +(\bar\eta^{(\rm
 rad)}\eta^{(\rm rad)}, ( - G_{0}^{-1} + {\cal M})^{-1})\right\}
 \eea
 with the effective action \cite{pre}
 \begin{eqnarray} \label{3-11}
W_{eff}[{\cal M}] = - \frac{1}{2}  ( {\cal M}, {\cal K}^{-1} {\cal
M} ) + i Tr \log  ( - G_{0}^{-1} + {\cal M}).
\end{eqnarray}
 The first step to the quantization of the effective action  is the
 determination of its minimum
 \begin{eqnarray}\label{8}
 \frac{\delta W_{eff} ({\cal M})}{{ \delta {\cal M}} }
  \equiv - {\cal K}^{-1} {\cal M} + \frac{i}{{ G_{0}^{-1}
 - {\cal M} } }|_{{\cal M}=\Sigma} = 0
 \end{eqnarray}
 known as the Schwinger -- Dyson equation of
 the fermion Green function
 \be\label{nak}
 \underline{G}_\Sigma(q)=\frac{1}{\rlap/ q-m_0-\underline{\Sigma}(q^{\bot})},
 \ee
 we denote the corresponding classical solution for the bilocal
 field by $ \Sigma (z)= \int d^{4}q \Sigma(q^{\bot})
e^{iq\cdot z)}\delta(z\cdot n)$.

 The next step is the expansion of the effective action around the
 point of minimum $ {\cal M} = \Sigma + \widetilde{{\cal M}}$,
 \begin{eqnarray}\label{9}
 W_{eff} ( \Sigma + \widetilde{{\cal M}} ) &= & W_{eff}^{(2)}+W_{int};
 \\\label{9a}
 W_{eff}^{(2)} (\widetilde{{\cal M}} ) &= & W_{Q}(\Sigma) + \left[ -
  \frac{1}{2} (\widetilde{{\cal M}} ,{\cal K}^{-1} \widetilde{{\cal M}}\right)
 + \frac{i}{2} \left( G_{\Sigma} \widetilde{{\cal M}} )^{2}\right ]
;  \\\label{9b}
 W_{int}=\sum_{n=3}^{\infty}W^{(n)}& = & i  \sum_{n=3}^{\infty}
 \frac{1}{n} ( G_{\Sigma} \widetilde{{\cal M}} )^{n}, \, \, \,
~~~~~~~~~  ( G_{\Sigma} = ( G_{0}^{-1} -
 \Sigma)^{-1} ),
 \end{eqnarray}
  where the small fluctuations of
the bilocal field $\widetilde{{\cal M}}(x,y)=
\widetilde{{\cal M}}(z=x-y|X=(x+y)/2)$ in the effective action (\ref{9})
 can be presented as the series of a complete set of the
 solutions $ \Gamma $ of the classical equation (known as
  the Bethe -- Salpeter one)
 \begin{eqnarray}\label{10}
 \frac{ \delta^{2}W_{eff} ( \Sigma + \widetilde{{\cal M}} )}{  { \delta
 \widetilde{{\cal M}}^{2}} } \vert_{ \widetilde{{\cal M}} = 0 }  \cdot \Gamma = 0
 \end{eqnarray}
 with a set of quantum numbers ($A$) including masses $M_A=\sqrt{{\cal
 P}_\mu^2}$ and energies $\omega_A=\sqrt{{\vec {\cal
 P}}^2+M_A^2}$.
 In the momentum representation this series takes the form
 \be\label{set}
 \widetilde{{\cal M}}(z|X)=\sum\limits_A\int\frac{d^3\vec {\cal
 P}}{(2\pi)^{3/2}\sqrt{2\omega_A}}
\frac{d^4q\, e^{iz\cdot q}}{(2\pi)^4}
 \{ e^{i {\cal P}\cdot {X}} \Gamma_A(q^{\bot}|{\cal P})a^+_A({\cal P})
 +e^{-i {\cal P}\cdot {X}} \bar{\Gamma}_A(q^{\bot}|-{\cal P})a^-_A({\cal P})\},
 \ee
 where $\widetilde{{\cal M}}(z|X)\sim \delta(z\cdot n)$ and $a^{+}_{A},a^{-}_{A'}$ are
the bound state creation and annihilation operators that obey
 the commutation relations
 \begin{eqnarray} \label{comrel}
 \biggl[
 a^{-}_{A'}( \vec{\cal P'} )     ,
 a^{+}_{A} ( \vec{\cal P}  )
 \biggr]   = \delta_{A'A}
 \delta^3 (
 \vec{\cal P'} - {\cal P} ) \,\,\, ,~~~
 \biggl[ a^{\pm}_A({\cal P}),a^{\pm}_{A'}({\cal P}')\biggl]=0~.
 \end{eqnarray}

  In the momentum representation
$\underline{\Gamma}(k^{\perp},{\cal P})$, Eq. (\ref{10}) takes the form
 \begin{eqnarray} \label{bs0}
 \underline{\Gamma}(k^{\perp}, {\cal P}) = i \int \frac{d^{4}q }{ (2\pi)^{4}}
 \underline {V} ( k^{\perp} - q^{\perp} ) \rlap/n \left[
 \underline{G}_{{\Sigma}a}(q+{ {\cal P} / 2 }) \Gamma(q^{\perp},
 {\cal P} ) \underline{G}_{{\Sigma}b}(q-{ {\cal P} / 2 }) \right]
 \rlap/n,
 \end{eqnarray}
  where $ G_{{\Sigma}a},~ G_{{\Sigma}b} $ are
   the fermion propagators   of the two particles ($a$)
 and ($b$).
Solutions of equation (\ref{10}) satisfy the normalization
 condition~\cite{Nakanishi}
 \be\label{nakanishi}
 i\frac{d}{d {\cal P}_0}\int \frac{d^4q}{(2\pi)^{4}}
 tr \left[  \underline{G}_\Sigma(q-\frac{{\cal
 P}}{2})\bar{\Gamma}_A(q^{\bot}|-{\cal P}) \underline{G}_\Sigma(q+\frac{{\cal
 P}}{2}){\Gamma}_A(q^{\bot}|{\cal P})
 \right]=2\omega_A
 \ee
The  corresponding Green function of the bilocal field takes form
 \be\label{green}
 {\cal G}(q^{\bot},p^{\bot}|{\cal P})=\sum\limits_H
 \left\{
 \frac{\Gamma_H(q^{\bot}|{\cal P})\bar{\Gamma}_H(p^{\bot}|-{\cal P})}
 {({\cal P}_0-\omega_H-i\varepsilon)2\omega_H}-
 \frac{\Gamma_H(q^{\bot}|-{\cal P})\bar{\Gamma}_H(p^{\bot}|{\cal P})}
 {({\cal P}_0-\omega_H-i\varepsilon)2\omega_H}~
 \right\}~.
 \ee
 The Bethe -- Salpeter  field
 $\Gamma(q^{\perp}, {\cal P})$ is
 connected with the  wave function
 by the relation~(see \cite{5,6})
\begin{eqnarray} \label{bs3}
 \Phi_{ab}(q^{\perp},{\cal P})&=&\frac{i}{2\pi }\int
 \left[\frac{{\cal P}\cdot dq} {M_A}
 \right]
  \left[
 \underline{G}_{{\Sigma}a}(q+{ {\cal P} / 2 }) \Gamma(q^{\perp},
 {\cal P} ) \underline{G}_{{\Sigma}b}(q-{ {\cal P} / 2 }) \right]
 \end{eqnarray}
 In the nonrelativistic limit in the frame $n=n^{\rm (cf)}=(1,0,0,0)$
this wave function reduces to the
 Schr\"odinger one (\ref{relative})
 \be\label{posi}
 \Phi_A^{\alpha\beta}(\vec q,M_A)=\left(\frac{1+\gamma_0}{2}
 \gamma_5\right)^{\alpha\beta}\underline{\Psi}_{Sch}(\vec q)
 \sqrt{\frac {M_A} 4}
  \ee
 satisfying  the standard Schr\"odinger equation
\be \label{relative}
 \left(-\frac {1}{2\mu}\frac {d^2}{d{\vec z}^2}
 +V(\vec z)\right)\underline{\Psi}_{Sch}(\vec z)=
 \epsilon\underline{\Psi}_{Sch}(\vec z)
 \ee
 where $\underline{\Psi}_{Sch}(\vec z)=
 \int\limits_{ }^{ }\frac{d^3q}{(2\pi)^3}e^{(i\vec q \vec z)}
 \Psi_{Sch}(\vec q)$ is a normalizable wave function
 ($\int d^3z{\parallel\underline{\Psi}_{Sch}(z)\parallel}^2=1$)
 in a comoving frame ${\cal P}_\mu=(M_A,0,0,0)$;
 here $M_A=(M_p+m_e-\epsilon)$ is the mass of an atom,
 $M_p,m_e$ are masses of proton and electron,
 $\mu = M_{p} \cdot m_{e} / ( M_{p}+m_{e})$ is the reduced mass, and
   $z_k= (x-y)_k$
  are   relative coordinates.

  \vspace{0.5cm}

{\bf A.2. S-matrix elements}

\vspace{0.51cm}

 The next step is the construction of
 the relativistic covariant S-matrix elements considered as
 probability amplitude transitions between the all asymptotic
 states. These states includes bound
 states  given by the spectral
 equation (\ref{16-5}) and all Lorentz transformations
 of the comoving frame distinguished by the time-axis $n^{\rm (cf)}$.
 The Lorentz transformations of the time-axis
 $n^{\rm (cf)}\to n^{\rm (ID)}$ mean the changes of the initial data (ID)
 of the bound states, i.e. their total momenta
${\cal P}_{A\mu} =M_An_\mu^{\rm (ID)}$. These bound states
 correspond to the
  instantaneous interaction with a new time-axis $n^{\rm (ID)}$
 \be\label{ID}
 {\cal K}( x,y \mid  n^{\rm (ID)}) =\rlap/ n^{\rm (ID)}
 V(z_\mu-n^{\rm (ID)}_\mu z \cdot  n^{\rm (ID)})
  \delta(z \cdot  n^{\rm (ID)} ) \rlap/ n^{\rm (ID)}.
\ee
 One can see that in the space of the bilocal fields (\ref{set})
 this unit time-axis $n^{\rm (ID)}$ is proportional to
 an eigenvalue of the
 total momentum ``operator'' (i.e. the derivative with respect to
total coordinate)
 \begin{eqnarray} \label{3-6}
  n^{\rm (ID)}_\mu  {\cal M} (z \vert X)   \sim
\frac{1}{i} \frac{\partial}{\partial X_\mu} {\cal M} (z\vert X)
 \end{eqnarray}
  in  agreement with the  Markov -- Yukawa definition
  of the relativistic simultaneity in terms of
  bilocal fields  ${\cal M}(z|X)\equiv {\cal M}(x,y)$ \cite{ym}
 \begin{eqnarray}\label{3-3}
 z_\mu \frac{\partial}{\partial X_\mu} {\cal M}(z \vert X) = 0~.
 \end{eqnarray}
  Recall that this  Markov -- Yukawa definition \cite{ym}
  of the relativistic simultaneity in terms of
 bilocal fields has a deeper mathematical
 meaning as a constraint of irreducible nonlocal
 representations of the Poincar\'e group \cite{wigner}
 for an arbitrary bilocal field
 $ {\cal M}(x,y) = {\cal M}(z \vert X)$~\cite{5,luc}.

 The  relativistic covariant constraint-shell
 quantization gives both
 the spectrum of bound states and their S-matrix elements.

 Using the decomposition over the bound state quantum numbers $(H)$
 \be\label{set1}
 \widetilde{\Phi}_{H(ab)}(q^{\bot}|{\cal P})=G_{\Sigma a}(q+{\cal P}/2)
 \Gamma_{H(ab)}(q^{\bot}|{\cal P}),
 \ee
 we can write the S-matrix elements for the interaction $W^{(n)}$~(\ref{9b})
 between the vacuum and  n bound states
 \be \label{S-matrix}
 <H_1{\cal P}_1, ...,H_n{\cal P}_n|iW^{(n)}|0>=-i(2\pi)^4 \delta^4
 \left( \sum\limits_{i=1 }^{n }{\cal P}_i\right) \prod\limits_{j=1 }^{n }
 \left[\frac{1}{(2\pi)^32\omega_j}\right]^{1/2}
 M^{(n)}({\cal P}_1,...,{\cal P}_n)
 \ee
 $$
 M^{(n)}=\int \frac{id^4q}{(2\pi)^4 n} \sum\limits_{\{i_k\} }^{ }
 \widetilde{\Phi}_{H_{i_{1}}}^{a_1,a_2}(q| {\cal P}_{i_1})
 \widetilde{\Phi}_{H_{i_{2}}}^{a_2,a_3}(q-\frac{{\cal P}_{i_1}+
{\cal P}_{i_2}}{2}| {\cal P}_{i_2})
 \widetilde{\Phi}_{H_{i_{3}}}^{a_3,a_4}\left(q-\frac{2{\cal P}_{i_2}+
 {\cal P}_{i_1}+{\cal P}_{i_3}}{2}| {\cal P}_{i_3}\right)
 $$
 $$
 ...\widetilde{\Phi}_{H_{i_{n}}}^{a_n,a_1}
 \left(q-\frac{2({\cal P}_{i_2}+...+{\cal P}_{i_{n-1}})+{\cal P}_{i_1}+
 {\cal P}_{i_n}
 }{2}| {\cal P}_{i_n}\right),
 $$
 where $\{i_k\}$ denotes permutations
 over $i_k$; the amplitude
 $<H_1{\cal P}_1, ...,H_l{\cal P}_l|iW^{(n)}|H_{l+1}
 {\cal P}_{l+1}, ...,H_n{\cal P}_n>$ can be obtained by the
 change of the momentum signs ${\cal P}_{l+1}, ...,{\cal P}_n \to {-\cal P}_{l+1}, ...,-{\cal P}_n$.

 Expressions~(\ref{set}),~(\ref{green}),~(\ref{set1}),~(\ref{S-matrix})
 represents Feynman rules for the construction of a quantum field theory
 with the action~(\ref{9}) in terms of bilocal fields.

 It was shown~\cite{a20} that this quantum field theory
 with a separable potential
 of the instantaneous interaction leads to
 the well-known Nambu Jona-Lasinio model~~\cite{a21,eb} and the
 phenomenological chiral Lagrangians~\cite{fpp} used for
 the description of the low-energy meson physics.

 In the context of the Dirac approach to gauge theory,
 to solve the problem of hadronization in QCD, one needs
 to answer the following questions:
 i) What is the origin of the ``separable potential'' of hadronization
 in the non-Abelian theory?
 ii) How to combine the Schr\"odinger equation for
 heavy quarkonia (that is derived by the residuum of poles
 of the quark Green functions) with the quark confinement?
 iii) What is the origin of the additional mass of
 the ninth pseudoscalar meson?

\section*{Appendix B: Instantaneous interactions in  QCD}

\renewcommand{\theequation}{B.\arabic{equation}}

\setcounter{equation}{0}


{\bf B.1. QCD action and monopole vacuum}

\vspace{0.51cm}


 QCD was proposed~\cite{fgl} as the non-Abelian $SU_c(3)$
 theory with the action functional
 \be \label{5u}
 W=\int d^4x
 \left\{-\frac{1}{4}{F^a_{\mu\nu}}^2
 + \bar\psi[i\gamma^\mu(\partial _\mu+{\hat
 A_\mu})
 -m]\psi\right\}~,
 \ee
 where $\psi$ and $\bar \psi$ are the fermion quark fields,
 ${\hat A_\mu}=g\frac{\lambda^a }{2i} A_\mu^a~$,
 \be \label{5v}
 F_{0i}^a = \partial_0 A^a_i - D_i^{ab}(A)A_0^b~,~~~~~~
 F_{jk}^a=\partial_jA_k^a-\partial_kA_j^a+g f^{abc}A^b_jA_k^c
 =\epsilon_{ijk}B_i^a
 \ee
 are non-Abelian gluon electric  and magnetic tensions, and
 \be\label{5cv}
 D^{ab}_i(A)=\delta^{ab}\partial_i + gf^{acb} A_i^c
 \ee
 is the covariant derivative.
 The action (\ref{5u}) is invariant with respect to
 gauge transformations
 $u(t,\vec x)$
 \be \label{gauge11}
 {\hat A}_{\mu}^u (x_0,\vec x)= u(x_0,\vec x)\left({\hat A}_{\mu} +
  \partial_\mu
 \right)u^{-1}(x_0,\vec x),~~~~~~
 \psi^u = u(x_0,\vec x)\psi~.
 \ee
  The class of functions with finite
 energy density  includes the monopole - type functions $A_i\sim
 O(1/r)$ like the Coulomb potential.

 The asymptotic freedom phenomenon  is considered
  as one of the arguments in favor of that the standard
    perturbation theory in QCD
  is unstable \cite{sh,ms}.
 One of candidates  of
  the stable vacua  in the Minkowskian space-time
  is the stationary  monopole - type solution
 of the classical equation
 \be\label{1su3}
 \hat A^{\rm vacuum}_0(x_0,\vec x)=0,~~~~\hat A^{\rm vacuum}_k(x_0,\vec x)=\hat
 \Phi_k(0,\vec x)= - i
 \frac{\lambda_A^a}{2}\epsilon_{iak}\frac{x^k}{r^2}f(r),
 \ee
 where the antisymmetric SU(3) matrices $\lambda_A^a$ are denoted by
 $
 \lambda_A^1:=\lambda^2,~\lambda_A^2:=\lambda^5,~\lambda_A^3:=\lambda^7
 $,
 and $r=|\vec x|$, the vacuum field contains only one function $f(r)$.

 In the following we discuss  the SU(2) case
   $\lambda_A^a \to \tau^a,  (a=1,2,3)$ \cite{6}. In this
  case,
 the classical equation for the function $f$ in Eq. (\ref{1su3}) takes the form
 \be\label{higgs}
 D_k^{ab}(\Phi_i) F_{kj}^b(\Phi_i)=0~~\Rightarrow~~\frac{d^2 f}{dr^2}+
 \frac{f(f^2-1)}{r^2}=0~.
 \ee
 We can see three solutions of this equation
 $
 f_1=0,~ f_{2}=+ 1,~f_{3}=- 1~~(r\not= 0)
  $.
 The first solution corresponds to the naive unstable perturbation
 theory. 
 Two nontrivial solutions  are the Wu-Yang monopoles \cite{wy}
 applied  to the construction of physical variables
 in~\cite{fn}.
 The Wu-Yang monopole is a solution of the classical
 equations everywhere besides the origin of the coordinates $r=0$.
 The corresponding magnetic field is
 $
 \label{sb} B_i^a(\Phi^{(0)}_k)={x^a x^i}/{(gr^4)}
 $. 
 Following Wu and Yang \cite{wy}, we consider the whole finite
 space volume, excluding  an $\epsilon$-region around the singular
 point. To remove
 a singularity at the origin of coordinates and regularize its
 energy, the Wu-Yang monopole is considered as  an infinite
 volume limit ($V_0  \to\infty$) of the
 Bogomol'nyi-Prasad-Sommerfield (BPS) monopole~\cite{BPS}
 \be
 \label{bps} f^{WY}={1}~\Longrightarrow~ f^{BPS}= \left[1 -
 \frac{r}{\epsilon \sinh(r/\epsilon)}\right]
 \ee
 with the finite energy
 $
  \int\limits_{
 }^{ }d^3x [B^a_i(\Phi_k)]^2 \equiv V_0 <B^2>
 ={4\pi}/{(g^2 \epsilon)}\equiv {1}/{(\alpha_s \epsilon)}
 $ 
 determining the finite volume  regularization
 of the monopole size
 $
\epsilon=[\alpha_s V_0<B^2>]^{-1}
 $. 
 The  vacuum energy-density of the monopole  solution $<B^2>$
 is removed by a finite counter-term in the
 Lagrangian.

 The color particle excitations ${\cal A}^{a(\rm rad)}_\mu$
\be\label{2su3}
 \hat A_0(x_0,\vec x)=\hat {\cal A}^{(\rm rad)}_0(x_0,\vec x),
 ~~~ \hat A_k(x_0,\vec x)=\hat
 \Phi_k(\vec x)+\hat {\cal A}^{(\rm rad)}_k(x_0,\vec x).
 \ee
 are considered
 as weak deviations in the vacuum background (\ref{1su3})
 by analogy with the Dirac radiation variables in QED
 (\ref{1c1}), (\ref{kc1}) with the constraint
$
 D_k^{ab}(\Phi_i) {\cal A}^{b(\rm rad)}_k(A)\equiv {0}
  $
 and the Gauss law
 \be\label{q1c1}
 \Delta^{ab}(\Phi) {\cal A}^{b(\rm rad)}_0=-j^{a(\rm rad)}_{t0},
  \ee
  where
 $
 \Delta^{ab}(\Phi)=[D^2_i(\Phi_k)]^{ab}
 $
 is the Laplacian in the vacuum background, and $j^{a(\rm rad)}_{t0}$ is
 the total color current.

 The  action
 can be expressed
 in terms of the  radiation - type variables
 \cite{6}
 \bea\label{c14-2}
 &&S_{\rm QCD}|_{\frac{\delta S_{\rm QCD}}{\delta A_0}=0}
 \equiv S^{(\rm rad)}_{\rm QCD}=\int d^4x
 \left[\frac{1}{2}A^{a(\rm rad)}_k{G^{ab}_{ki}}^{-1}A^{b(\rm rad)}_i+
 A^{(\rm rad)}_kj^{(\rm rad)}_{k}\right]\\
 \nonumber
  &&+(\psi^{(\rm rad)
 }\bar \psi^{(\rm rad},G_m)-\frac{1}{2}\int dx^0d^3xd^3y
 j_{t0}^{a(\rm rad)}(x)V^{ab}(x-y)j_{t0}^{b(\rm rad)}(y)+
 ...+S_{\rm soc.}(\eta^{(\rm rad},\bar \eta^{(\rm rad}),
 \eea
 where
 \be\label{soc-r1}
 S_{\rm soc.}(\eta^{(\rm rad},\bar \eta^{(\rm rad})=\int d^4x \left[
 \bar \eta^{(\rm rad)}\psi^{(\rm rad)}+
 \bar\psi^{(\rm rad)}\eta^{(\rm rad)} \right]
 \ee
 is the source action,
 $G^{ab}_{ki}$ and $G_m$ are the Green functions of the
 gluon and quark in the background field $\Phi^a_k$,
 $$
 V^{ab}({\vec z})=  -\frac{z^a(z)z^b(z)}{|\vec z|^2}\frac{1}{4\pi|\vec z|} +
 \sum\limits_{\alpha=1,2} e^a_{\alpha}(z)
 e^b_{\alpha}(z)\left( d_- |{\vec z}|^{l_-} +
 d_+ |{\vec z}|^{l_+}\right)
 $$
 is the  potential \cite{6,bpr} considered as the Green function
\be\label{q4c}
 [\Delta^{ab}(\Phi(\vec z))]V^{bc}(\vec z)
 =\delta^{ac}\delta^3(\vec z);
 \ee
 here $\vec z={\vec x}-{\vec y}$, $e^a_{\alpha}(z)$ are
 a  set of orthogonal vectors in color space
 $z_a e^a_{\alpha}(z)=0$, $d_{\pm}$ are constants,
 and ${l_+},~{l_-}$ can be
 found as  roots of the ``gold section'' equation $l^2 + l =1$:
 \begin{equation} \label{fun}
 {l_+} =-\frac{1+\sqrt{5}}{2}\approx - 1.618~
 ;~~~{l_-} =\frac{-1+\sqrt{5}}{2}\approx 0.618~.
 \end{equation}
 The solution of the set of Schwinger -- Dyson and Bethe --
 Salpeter
 equations for the rising potential
 was considered in  numerous papers, where
 the effect of spontaneous
 chiral symmetry breaking was described (see ~\cite{5,6,puz}).
 Therefore, this Green function contains the rising
 potential as the origin of
 the 'hadronization' of quarks and gluons
 in QCD \cite{puz}.

 The Wigner -- Markov -- Yukawa relativistic generalization
 of the instantaneous interaction in this case can lead to
 the bilocal
 theory with the spontaneous chiral symmetry breaking
 in the low-energy limit with
 the effective chiral Lagrangian one \cite{vp1,gs}.


 \vspace{.5cm}

{\bf B.2. Zero mode of Gauss' constraint}

\vspace{0.51cm}

 The BPS - type regularization (\ref{bps}) of the monopole vacuum
 is interesting by the existence of
  a nonzero solution of homogeneous equation
\be\label{q4}
 [\Delta^{ab}(\Phi)]{\cal Z}^a=0
 \ee
 of the type of  the  BPS solution~\cite{BPS}
 \be \label{pgc0}
 \hat {\cal Z}(\vec x)=-
 i \pi\frac{\tau^a x^a}{r} f_0^{BPS}(r)~,~~~~
 f_0^{BPS}(r)=\left[
 \frac{1}{\tanh(r/\epsilon)}-\frac{\epsilon}{r}\right]~.
 \ee
 The complete solution of the Gauss equation  (\ref{q1c1}), in this case, is
 the sum of a solution of the homogeneous equation (\ref{q4})
 and a solution of the nonhomogeneous equation:
 \be\label{q3}
 {\cal A}^{a(\rm rad)}_0(\vec x)=c(x_0){\cal Z}^a
 - \int d^3y V^{bc}(\vec x -\vec y)j^{b(\rm rad)}_{0}(\vec y),
  \ee
 where $c(x_0)$ is the zero mode variable.

 The BPS vacuum model gives  the possibility to compare
 the radiation variables in the action (\ref{c14-2}) with the
   set of the Dirac - type  variables
   removing the  field $A_0$ by the gauge transformations
 \bea
 \nonumber
 \Phi_k^{(\rm D)}&=&U^{(\rm D)}
 [\Phi_k^{(\rm rad)}+\partial_k]{U^{(\rm D)}}^{-1},\\
 \label{c13-1}
 {\cal A}^{(\rm D)}_l&=&U^{(\rm D)}
 [{\cal A}_k^{(\rm rad)}]{U^{(\rm D)}}^{-1}
 ,\\
 \nonumber
 \psi^{(\rm D)}&=&U^{(\rm D)}\psi^{(\rm rad)},
 \eea
 where
 \be\label{N1}U^{(\rm D)}=\exp\left\{\int^{x^0} d \bar x^0 c(\bar x^0)
  \hat Z(\vec
 x)\right\}\equiv \exp\{N (x^0) \hat Z(\vec
 x)\}.\ee
 The Dirac variables (\ref{c13-1})
 ({\it dressed} by the zero mode phase factor)
 differ from the {\it radiation}
 ones  in the action (\ref{c14-2}). The differences are
 Abelian U(1) anomaly and
 the new {\it gauge of physical sources}
\be\label{soc-d1}
 S_{\rm soc.}(\eta^{(\rm D},\bar \eta^{(\rm D})=\int d^4x \left[
 \bar \eta^{(\rm D)}\psi^{(\rm D)}+\bar\psi^{(\rm D)}\eta^{(\rm D)}
 \right]
 \ee
  as consequences of  the gauge transformations (\ref{c13-1}).

The zero mode of the Gauss constraint (\ref{q4})
 leads to the vacuum electric tension \cite{p2,gip}
 \be \label{35-1}
 {F^{a}_{0i}}_{\rm vac}= c(x_0) D_i^{ab}({\Phi}){\cal Z}^b
 \equiv\dot N~D_i^{ab}({\Phi}){\cal Z}^b~~~~~~~~~~
 (\dot N =\partial_0 N).
 \ee
 The BPS monopole electric tension is
  proportional to the vacuum magnetic one. If we choose
  the coefficient of the proportionality as
 \be\label{qsoc-d1}
 D_i^{ab}({\Phi}){\cal Z}^b=B^a_i(\Phi)
 \frac{\alpha^2_s}{(2\pi)^2} \frac{1}{V_0 <B^2>} ~~~~~
 \left(\frac{g^2}{4 \pi}\equiv\alpha_s\right),
 \ee
 the zero mode
  $c(x_0)=\dot N$
  can be identified  with
  a variable  given by the winding number functional
 $$\nu=\frac{g^2}{16\pi^2}\int\limits_{x^0_{in} }^{x^0_{out}
 }dx^0
 \int d^3x F^a_{\mu\nu} \widetilde{F}^{a\mu \nu}=\frac{\alpha_s}{2\pi}
 \int d^3x D_i^{ab}({\Phi}){\cal Z}^bB_i^b(\Phi)
 \int\limits_{x^0_{\rm in} }^{x^0_{\rm out} }dx^0\dot N
 =N(x^0_{\rm out}) -N(x^0_{\rm in})~ $$
 showing that $N$ can interpolate between different classical vacua
 like an instanton \cite{ins}\footnote{In this case, the monopole
 interpolation amplitude $\exp\{iP_N[N(x^0_{\rm out})-N(x^0_{\rm in})]\}$
 coincides with the instanton one
 $\exp\{-\dfrac{8\pi^2}{g^2}(n_{\rm out}-n_{\rm in})\}$ for a
 nonphysical value of the momentum $P_N=i\dfrac{8\pi^2}{g^2}$.}
 \be \label{35-2}
 N(x^0_{\rm out})=n_{\rm out}, ~~N(x^0_{\rm in})=n_{\rm in}.
 \ee
 The equations (\ref{35-1}) and  (\ref{qsoc-d1}) lead to   the vacuum action
  of the zero mode dynamics
 \bea\label{35-4a}
  W_{\rm QCD} &=&\int dx^0  \int\limits_{V_0 }^{ }d^3x \frac{1}{2}
  ({F^{a}_{0i}}_{\rm vac})^2=
 \int dx^0 \frac{\dot N^2}{2}
 \left(\frac{2\pi}{\alpha_s}\right)^2\frac{1}{V_0<B^2>}~.
 \eea
 As we pointed out above, one of consequences
 of the {\it dressed} variables (\ref{c13-1}) is the appearance
 of the Abelian anomaly
  \cite{eb}
 in the effective meson action obtained by the bosonization
 of QCD of the type of (\ref{9})
 \be \label{sm}
 W_{eff}=\int dx^0 \left\{
 \frac 1 2
 \left({\dot\eta_0}^2-m_0^2\eta_0^2\right)
 V_0 +
 C_\eta\eta_0 \dot N \right\}
 \ee
 in the pseudoscalar $\eta_0$ meson isosinglet channel, where
 $
 C_{\eta}=\dfrac{N_f}{F_\pi}\sqrt{\dfrac 2 \pi},
 N_f=3$ is the number of flavors, and
 $F_\pi=92$ MeV is the weak decay coupling.
 This action is known as the  Veneziano one \cite{2}.

 The diagonalization of the total Lagrangian
 supplemented by the action (\ref{35-4a})
leads to an additional mass
 of the pseudoscalar  $\eta_0$ meson
 $
  \triangle {m_\eta}^2
 = \dfrac{N_f^2}{F_{\pi}^2}\dfrac{\alpha_s^2<B^2>}{2\pi^2}
 $ \cite{6}.
 This result allows us to estimate the value of the vacuum chromomagnetic
 field  in QCD \cite{bl}
 $$
 <B^2>=\frac{}{}\frac{2\pi^3F_{\pi}^2\triangle
 {m_\eta}^2}{N_f^2\alpha_s^2}=\frac{0.06 GeV^4}{\alpha_s^2}~.
 $$
 After calculation
 we can remove infrared  regularization $V_0 \to \infty $.

  Thus, there are observational arguments in favor of
  that physical fields are identified with
  the {\it dressed} one  (\ref{c13-1}). However,
   the gauge of {\it dressed} sources  (\ref{soc-d1}) differ
 from the {\it radiation}
  ones (\ref{soc-r1}) by the  phase factors (\ref{N1}) depending
on the winding number variable $N$.
     The initial position of winding number variable $N$ are
    degenerated like the U(1) initial data $e^{ix_{\rm(in)}}=1, x_{\rm(in)}=2\pi k$,
where $k$ is an integer number.
 The  phase factors (\ref{N1})
 are determined within the matrixes
$v^{(n)}(\vec x)=e^{n\hat {\cal Z}(\vec{x})}$, where $n$ is an integer number
 which counts how many times a three-dimensional path $v^{(n)}(\vec{x})$
 turns around the $SU(3)$-manifold when the coordinate $x_i$ runs
 over the space where it is defined.
 Even, if we choose  the zero electric monopole
   energy   fixing the zero momentum $P_N=0,N_{\rm (in)}=0$,
    we do not know a position of winding number variable due to
   this topological degeneration
 of all physical states
    with respect to the phase factor (\ref{c13-1})
   $U^{(D)}=e^{n\hat {\cal Z}(\vec{x})},
  n=0, \pm 1, \pm 2, ...$.
   Therefore, the creation amplitude of the {\it dressed} color
 particles $\psi^{(D)}=\psi^{(\rm rad)}U^D$
   should be averaged over all degenerated states
   \be\label{conf1}
   \frac{1}{2L}\sum\limits_{n=-L}^{n=+L}<\mbox{vacuum}|\psi^{(\rm rad)}U^D|q>=
   \frac{e^{i\omega x_{(\rm in)}^0}}{(2\pi)^{3/2}
   \sqrt{2\omega}}
\frac{1}{2L}\sum\limits_{n=-L}^{n=+L} e^{n\hat{\cal Z}(\vec{x})}=0,
   \ee
 where $|q>$ is the Dirac  one-particle state (\ref{c13-1})
with a quantum
 number $q$.
 Averaging over all parameters of the topological initial data
 can lead to a complete destructive interference of all color
 amplitudes~\cite{p2} and
 the  zero probability amplitude of creation of
  color quarks or gluons; whereas
  their {\it radiation} propagators
 have poles like the propagators of electron and photons and
 the colorless amplitudes of the type of the current production
$<j_\mu^{Da}j_\nu^{Da}>$=$<j_\mu^{{\rm (rad)}a}j_\nu^{{\rm (rad)}a}>$
 do not disappear.
 In this case, only colorless
 (``hadron'') states have to form a complete set of physical states.

 Disappearance of color  physical states
 due to the topological degeneration (\ref{conf1}) can be considered
 as  color confinement,
  compatible with the quark -- hadron duality  as the accepted method
 of analysis of observational data \cite{feynman}.



} {

\end{document}